# Adaptive control of dynamic networks

Chunyu Pan[1,2,3], Xizhe Zhang[1,4]*, Haoyu Zheng[4], Yan Zhang[4], Zhao Su[4], Changsheng Zhang[2], Weixiong Zhang[3,5,6]

1. Early Intervention Unit, Department of Psychiatry, Affiliated Nanjing Brain Hospital, Nanjing Medical University, Nanjing, China;
2. Northeastern University, Shenyang 110169, China;
3. Department of Health Technology and Informatics, The Hong Kong Polytechnic University, Hong Kong;
4. School of Biomedical Engineering and Informatics, Nanjing Medical University, Nanjing, Jiangsu 210001, China;
5. Department of Computing, The Hong Kong Polytechnic University, Hong Kong;
6. Data Science and Artificial Intelligence, The Hong Kong Polytechnic University, Hong Kong;

*Correspondence: Xizhe Zhang: zhangxizhe@njmu.edu.cn;

***Abstract*****—**Real-world network systems are inherently dynamic, with network topologies undergoing continuous changes over time. Previous works often focus on static networks or rely on complete prior knowledge of evolving topologies, whereas real-world networks typically undergo stochastic structural changes that are difficult to predict in advance. To address this challenge, we define the adaptive control problem and propose an adaptive control algorithm to reduce the extra control cost caused by driver node switching. We introduce a node-level adaptive control metric to capture both the stability and consistency of each node across historical topologies. By integrating this metric with a partial matching repair strategy, our algorithm adjusts the minimum driver node set in real time at each snapshot, while minimizing unnecessary reconfigurations between consecutive time steps. Extensive experiments on synthetic and real-world dynamic networks demonstrate that the proposed adaptive control algorithm significantly outperforms the existing algorithm, reducing the switching cost by an average of 22% in synthetic networks and 19\% in real-world networks, without requiring foreknowledge of the future evolution of the network. These findings extend the theoretical scope of dynamic network controllability and open new avenues for practical applications in transportation, social, and molecular regulatory systems.

## I. INTRODUCTION

Complex systems are common in many fields and are often represented as networks. They are essential for understanding the fundamental properties of these systems [1], [2], [3]. For instance, in the biomedical field, analysis of gene regulation and protein–protein interactions within cells [4], [5], [6], [7] provides a paradigm for discovering and interpreting, molecular pathways and disease mechanisms. In neuroscience, investigating control principles of brain networks [8], [9], [10], [11], [12], [13] can help elucidate the underlying mechanisms of brain function and cognition [14], as well as the origins of neurodegenerative diseases and neuropsychiatric disorders. In economics, the analysis of financial networks [15], [16], [17] facilitates risk assessment and economic trend prediction.

Network control has attracted much attention lately [18], [19], [20]. A network is considered controllable if external control signals can drive it from any initial state to any desired f inal state, with the nodes receiving control signals referred to as driver nodes [18], [21]. The set with the minimum number of driver nodes is called the Minimum Driver node Set (MDS). Liu and Barab´asi were among the first to explore the theoretical underpinnings of MDS [18], [21], which laid the foundation for subsequent studies on MDS properties and applications. Based on this, Ruths categorized driver nodes into three distinct types to elucidate network control structures and help unveil the essential roles of driver nodes and MDS compositions [22], [23]. However, the MDS in any given network is often not unique, and computing all possible MDS configurations is an NP-hard problem [24], presenting significant challenges. To address this issue, we previously proposed an efficient algorithm [25], [26] to identify all possi ble driver nodes by finding one maximum matching of a given network, avoiding the computational difficulties mentioned above. We also presented earlier a novel method [27], [28] for finding preferential matchings to derive MDS with distinct attributes, showing that reversing or removing edges can shift the network's control modes [29], [30]. D'Souza et al. [19] comprehensively reviewed network control and discussed its inception, the latest development and future research direc tions. These previous works have set a foundation for selecting optimal MDS.

Despite these achievements, it is important to highlight that most of these studies predominantly considered static networks with fixed topologies. However, most real-world networks are inherently dynamic [31], [32], with topologies evolving constantly, seriously limiting control theory's applicability in practice. As a result, increasing attention has been devoted to the controllability of dynamic and temporal networks. A common approach is to treat a dynamic network as a sequence of static snapshots and identify a global or time-invariant MDS. For example, Posfai et al. [33] proposed a subspace controllability model under known topological evolution. Ra vandi et al. [34] extended this to a heuristic method, and Yao et al. [35] further developed controller-switching strategies based on temporal structures. These studies provide important theoretical foundations for dynamic network control.

An implicit but critical assumption underlying most of the above methods is that the entire temporal evolution of the network is known a priori. That is, they treat the dynamic network as a fully observable object and aim to find a static or global MDS that applies across all time steps. This assumption is often violated in practice—a scenario that may occur only in specific recurrent dynamic networks, such as artificially designed ones. However, the structures of most real-world networks vary stochastically. Real-world networks—such as social, transportation, and communication systems—typically evolve in stochastic and unpredictable ways. When future network changes are unpredictable, precomputing an MDS suitable for the entire dynamic network is impractical.

This leads to a fundamentally different problem setting: at each time step, a controller must make decisions based solely on the current and past network structure, with no access to future information. Therefore, it is imperative to update the MDS dynamically as the network evolves, especially in the absence of prior knowledge about future structural changes. In this context, adaptive control of dynamic networks emerges as a new challenge—how to construct a time series of MDSs that not only ensure controllability at each step, but also maintain temporal consistency to minimize control cost and structural disruptions. This online, history-dependent control paradigm has rarely been systematically explored and forms the core focus of this study.

Figure 1 illustrates the fundamental structure of this decision-making problem. At each point in time, control strategies must be determined based solely on the current and previous network states, without access to future information. A central challenge in this context is how to ensure effective control at each step while also maintaining stability in the control configuration over time. This naturally leads to a new class of control problems, in which the MDSs selected at different time points may vary significantly, resulting in extra control cost due to frequent reconfiguration.

Here, we present a novel control concept, termed adaptive control, to address the real-time decision-making challenge in dynamic networks. Unlike traditional approaches that rely on global knowledge of the network's temporal evolution [33], [34], [36], [35], adaptive control does not aim to manage the entire dynamic network in a post hoc manner. Instead, it seeks to ensure real-time controllability of each network snapshot as it emerges. When topological changes occur, the adaptive control strategy reconstructs a new set of driver nodes to maintain the controllability of the evolving system. To achieve this, we introduce a node-level control importance metric that leverages historical topological information and local structural cues to guide the selection of driver nodes. This metric is integrated into a partial matching repair algorithm designed to produce a new driver node set that remains as consistent as possible with its predecessor, thereby reducing unnecessary variations in the control structure over time.

In summary, we make the following contributions:

(1) We propose an adaptive control problem aimed at dynamically adjusting the MDS in real time to maintain control of the network as its topology changes. From the perspective of adaptive control, there is no need for prior knowledge of all network topology changes followed by post analysis. However, adaptive control faces a key challenge: the adjusted MDS may differ significantly from the previous MDS, introducing additional control costs.

(2) We present a new Adaptive Control algorithm (AC) to tackle the fundamental challenges of adaptive control in dynamic networks. The AC algorithm aims to minimize the difference between consecutive MDSs used in the adaptive control of dynamic networks. It selects the most suitable MDS for every consecutive network snapshot, considering its histor ical topological variations and potential future configurations.

The remainder of the paper is structured as follows: Section II introduces the related work of this paper. Section III defines dynamic networks and describes adaptive control. Section IV motivates the research and describes the AC algorithm. Section V presents empirical results on both synthetic and real-world networks, followed by an extended comparative analysis with multiple baseline methods. Finally, Section VI concludes and outlines future directions.

Fig. 1. Adaptive control of dynamic networks. At the current time tn, only the previous network topologies [t1,..., tn] are available, while the subsequent network topology is unknown. The goal of adaptive control is to compute the current MDSn in real time to maintain control of the network. As a result, each snapshot will have its own MDS.

## II. RELATED WORKS

Controlling complex networks has attracted extensive re search attention in recent years. Structural controllability pro vides a theoretical basis for understanding how a network's topology influences its controllability. In seminal work, Liu et al. [18] showed that the minimum set of driver nodes required to control a static directed network can be identified via a maximum matching on the network's bipartite repre sentation. This foundational result spurred a wide range of follow-up studies on network controllability and driver node selection. For example, various strategies have been proposed to minimize the number of driver nodes or alter network structure to enhance controllability in static networks [22], [37]. However, these works assume fixed topologies and do not address networks where connectivity changes over time.

Many real-world systems are dynamic networks with time varying topologies. Early efforts to extend structural control lability to temporal settings treated a dynamic network as a sequence of static "snapshot" graphs. P´osfai and H¨ovel [33] provided an early analytical framework for time-varying networks, examining how a single control input can steer a temporal network. They introduced a layered network model to study the controllable subspace under switching topologies, and discovered a phase transition in the size of the controllable subnetwork as the frequency of topology changes increases. Around the same time, Pan and Li [36] introduced the concept of controlling centrality to quantify the ability of individual nodes to control a temporal network. They developed graphical tools to classify temporal network structures and derived bounds on each node's controlling centrality, revealing that nodes with higher aggregated degree tend to have greater influence in controlling time-varying networks. These studies [33], [36] highlighted that temporal dynamics can actually aid controllability– for instance, certain networks that are uncontrollable in a static sense can be controlled when their interactions are properly sequenced in time. In the latest re search, Tu et al. [38] proposed a graphical criterion to evaluate the controllable subspace of temporal networks. Nonetheless, these efforts primarily focused on understanding controllability characteristics rather than designing online control strategies.

Subsequent research began investigating control strategies tailored to dynamic networks. An important line of work con sidered a moving or switching control input that can re-target different nodes over time. Qin et al. [39] proposed a method to restore controllability in temporal networks by predicting link failures. Their approach leverages network embedding and feature extraction to anticipate missing connections and reconfigure the network accordingly. This solution is specifi cally tailored to dynamic and evolving topologies. Yao et al. [35] proposed a single switching controller approach, wherein one control signal is dynamically relocated across network nodes to maximize influence within a given time window. They devised several switching strategies and demonstrated that a carefully timed moving controller can drive more nodes to desired states than any fixed controller placement. While effective in improving reachability, this approach assumes only one control input and requires predefining or computing an optimal switching schedule through the entire temporal duration. Other studies leveraged temporal variation to im prove control with multiple inputs. For instance, Cui et al. [32] demonstrated that properly segmenting the timeline by activating different links during different intervals can reduce the required number of controllers, a technique referred to as temporal segmentation. By scheduling the activation of edges, their method can make an otherwise uncontrollable network structurally controllable, illustrating that timing of connections can serve as an additional control lever. However, such methods generally operate offline– either assuming the temporal sequence is known in advance [35], or optimizing control decisions in predetermined time slots [32]. They do not explicitly address how to adapt control actions on-the-fly when the network evolution is unpredictable.

More recent efforts have started to explore online control in evolving networks. Some researchers formulated the driver node selection over time as a combinatorial optimization prob lem. For instance, Srighakollapu et al. [40] derived conditions for the structural controllability of temporal networks and proposed submodular optimization-based algorithms to iden tify driver nodes that ensure controllability over time. Their approach constructs a time-unfolded network and selects input nodes to maximize the dimension of the reachable subspace across the temporal horizon, but the full time-expanded graph still needs to be rebuilt when the underlying network structure changes. Importantly, their method—like most existing ap proaches—relies on the assumption that the complete temporal topology is known in advance, which limits its applicability in real-time or online settings where future configurations are unavailable. Similarly, Ravandi et al. [41] introduced a heuristic method for efficiently identifying a near-minimal set of driver nodes in temporal networks. By repairing a previous reachable subspace path sets solution, their algorithm finds an approximate MDS for the new snapshot more efficiently than a full re-calculation. Qin et al. [42] introduced the Online Temporal Acceleration Heuristic Algorithm (OTaHa), which significantly improves the efficiency of driver node detection in temporal networks. Unlike prior methods, OTaHa not only provides high accuracy but also reduces execution time by up to 99%, making it suitable for large-scale temporal networks. In recent advancements, Li [43] introduced a fully dynamic controllability method for temporal networks, which takes all network dynamics, such as node and link additions or removals, into account. Their method, implemented with a Controllable Dynamics Temporal Network (CDTN) model, significantly improves the efficiency of the controllability process by reducing overhead and computational complexity. The proposed polynomial-time algorithm for finding MDSs outperforms traditional methods in terms of speed, overhead, and the size of MDS. These strategies improve efficiency but are also designed under the assumption of known global temporal structure, and do not explicitly optimize the similarity between driver sets across time steps, potentially leading to large variations in the control configuration over time.

In fact, most existing dynamic network control algorithms focus on ensuring controllability over the entire temporal span or improving the efficiency of driver node selection. However, they seldom consider the temporal consistency of control inputs, which can lead to abrupt and frequent changes in the driver node set over time. Such fluctuations incur high implementation costs and may disrupt the system. This issue has only recently received attention and remains largely underexplored, particularly in online scenarios where control decisions must be made sequentially without knowledge of future topologies.

In summary, prior work has laid a foundation for under standing and controlling time-varying networks. However, to the best of our knowledge, no existing approach guarantees continuous, adaptive controllability in a dynamic network while also minimizing the temporal variability of the control inputs. This leaves a gap in the literature regarding truly online algorithms that can maintain control of a network in real time without prior information and with minimal reconfiguration at each change. Our work addresses this gap by introducing an adaptive control framework that constructs a time series of driver node sets to keep the system controllable at every step, while ensuring successive sets are as consistent as possible. Table I summarizes representative related works in context.

**Table 1**. Summary of related works

| Ref. | Contributions | Specificities |
|---|---|---|
| [18] | Established structural control lability via maximum matching in static networks. | Assumes fixed topology; not suited for real-time or dynamic scenarios. |
| [37] | Studied bimodal control modes in complex networks. | Not applicable to evolving systems; focuses on structural ef fects. |
| [22] | Defined control profiles to characterize node roles. | Static-only; no temporal net work modeling. |
| [33] | Layered graph model for tem poral controllability. | Requires known topology; offline; analyzed under single input assumption. |
| [36] | Proposed controlling centrality in dynamic settings. | Non-algorithmic; needs full temporal structure. |
| [38] | Proposed a graphical criterion to evaluate the controllable subspace of temporal networks. | Using maximum matching in temporal networks. |
| [39] | Proposed a method for restor ing controllability in temporal networks. | Focuses on dynamic link re covery. |
| [35] | Controller switching heuristics for reachability. | Predefined control schedule; single input. |
| [32] | Temporal segmentation to im prove controllability. | Offline only; assumes future edge activity known. |
| [40] | Submodular optimization for driver node selection. | Requires full time-expanded graph; not online-compatible. |
| [41] | Heuristic repair for near optimal driver sets. | Temporal inconsistency not ad dressed; assumes known evolu tion. |
| [42] | Proposed the OTaHa algorithm based on submodular optimization. | Reduces execution time by 99%; requires all topologies to be known. |
| [43] | Proposed a fully dynamic controllability method considering all temporal dynamics. | Reduces overhead and im proves control efficiency; requires all topologies to be known. |

## III. Problem Definition and Preliminaries

### *A. Controllability of dynamic network*

A dynamic network can typically be modeled as a linear time-varying (LTV) system, and following [44], its dynamics can be represented as

$$\frac{d\boldsymbol{x}(t)}{dt} = \boldsymbol{A}(t)\boldsymbol{x}(t) + \boldsymbol{B}(t)\boldsymbol{u}(t)$$

where state vector $\boldsymbol{x}(t) = (x_1(t), \dots, x_{N_t}(t))^{\mathbf{T}} \in \mathbb{R}^{N_t}$ represents the states of all $N_t$ nodes at time $t$, with $x_{\mathrm{n}}(t)$ denoting the state of node $n$. The number of nodes $N_t$ is not necessarily fixed but can change at certain time instants. $\boldsymbol{A}(t) \in \mathbb{R}^{N_t \times N_t}$ denotes the adjacency matrix of the dynamic network at $t$, $\boldsymbol{B}(t) \in \mathbb{R}^{N_t \times M_t}$, where $M_t \leq N_t$ specifies the nodes to be controlled by the external controller at $t$, and $\boldsymbol{u}(t) = (u_1(t), \dots, u_{M_t}(t))^{\mathbf{T}} \in \mathbb{R}^{M_t}$ defines the input on $M_t$ nodes. The controller uses input $\boldsymbol{u}(t)$ to control the system, where each control signal $u_{\mathrm{m}}(t)$ can typically influence multiple nodes, depending on the system's control structure. The set of nodes controlled by the external signals is determined by the input matrix $\boldsymbol{B}(t)$, which maps the control signals to the corresponding network nodes.

The interactions in most real-world systems occur at dis crete time points, such as email timestamps and human interactions [45], while the system dynamics remain con tinuous within each interval. This observation supports the snapshot-based modeling approach. The interactions occurring within time windows of duration τ, spanning the interval [t, t + τ), are aggregated. This models dynamic networks **G** as sequences of evolving snapshots [46], represented as **G**={$G_1(V_1,E_1)$,$G_2(V_2,E_2)$,...,$G_T(V_T,E_T)$,...}, where $G_i$ =($V_i$,$E_i$) denotes the i-th snapshot with node set $V_i$ and edge set $E_i$. This flexible modeling reflects the nature of many real-world systems, where the active node and edge set may change over time. The system is assumed to spend a finite amount of time in each snapshot [47], [48], and to switch instantaneously to the next one $G_{i+1}$.

From this point onward, the temporal evolution of the system is described with respect to the snapshot index *i*. The snapshot $G_i$, referred to as a temporal subgraph, representing the network structure at "time" *i*. Within temporal subgraph $G_i$, the topology is fixed as **A**(t) = $\mathbf{A}_i$ and **B**(t) = $\mathbf{B}_i$, and the number of nodes remains constant as $N_i = |V_i|$. In contrast, the state vector **x**(t) continues to evolve continuously under these fixed parameters. This assumption does not yield a discrete-time LTV system: only the topology switches at discrete instants, while the system state remains continuous. At the current time ϵ, only {$G_1$,...,$G_\epsilon$} are known, whereas future topological changes remain unknown (Figure 1).

Under the above assumptions, the dynamic system can be viewed as a sequence of linear time-invariant (LTI) systems [31]. For the i-th temporal subgraph $G_i$, the state vector is x(t) ∈ $\mathbb{R}^{N_i}$, and its dynamics are given by

$$\frac{d\boldsymbol{x}(t)}{dt} = \boldsymbol{A}_i\boldsymbol{x}(t) + \boldsymbol{B}_i\boldsymbol{u}(t)$$

According to the Kalman rank condition [49], $G_i$ is controllable if and only if the matrix $C_i = (B_i, A_iB_i, A_i^2B_i, \dots, A_i^{N_i-1}B_i)$ is full rank, and the system can be driven from any initial state to any final state in finite time. However, for most real networks, it is nontrivial to obtain the weight for every edge. To circumvent this difficulty, *Liu* [18] introduced structural controllability to assess network controllability only based on network structures. It provides an operational means to determine the minimum number of inputs required to control a network and identify the locations in the network to place the inputs.

In the structural controllability theory, the MDS signifies the smallest node assembly needed to guide a directed network to a desired state. The computation of the MDS is achieved by f inding a maximum matching in the bipartite graph of the network [18]. A matching in a network refers to a set of edges sharing no node in common. A maximum matching is a matching with the largest number of edges, which can be found by the Hopcroft-Karp (HK) algorithm [50]. Nodes without directly matched edges pointed to are designated driver nodes [18].

### *B. Adaptive control of dynamic networks*

Consider a dynamic network **G** with a series of temporal subgraphs {$G_1(V_1,E_1)$,$G_2(V_2,E_2)$,...,$G_T(V_T,E_T)$,...}, where $G_i(V_i,E_i)$ is the i-th temporal subgraph at time i. The objective of adaptive control is to derive an MDS sequence, denoted as DS =

$\{D_1, D_2, \ldots, D_T, \ldots\}$, which we refer to as the Driver set Sequence. Here $D_i$ represents the MDS of temporal subgraph $G_i$ at time i.

It is worth noting that $D_i$ and $D_{i-1}$ may differ significantly, which can lead to extra control costs. Therefore we need to ensure that the newly computed MDS is as similar as possible to the previous MDS to impose minimal disruption to the control scheme. To this end, we introduce the Extra Control Cost (ECC) to quantify the difference between consecutive MDSs of all consecutive temporal subgraphs:

$$ECC = \sum_{i=2}^{T} |D_i - D_{i-1}|$$

where $D_i$ represents the MDS of the i-th temporal subgraph, and $|D_i - D_{i-1}|$ is the number of driver nodes belonging to the current MDS but not to the previous one. In other words, ECC quantifies the number of new driver nodes required to adapt to the network changes. Our objective is to minimize ECC to reduce overall extra control costs:

Objective:

$$\text{Minimize } ECC(DS)$$

Subject to:

$$s.t.\begin{cases} DS = \{D_1, D_2, \ldots D_i, \ldots\} \\ \forall i, D_i \in All\ possbile\ MDS\ of\ G_i(V_i, E_i) \end{cases}$$

It is important to note that minimizing the ECC over a tem poral subgraph sequence constitutes an online combinatorial optimization problem, which is inherently non-convex and NP hard, as it involves selecting different MDSs from multiple valid maximum matchings at each snapshot. Therefore, we do not aim to achieve global ECC optimality, but instead focus on practically minimizing ECC through heuristic methods.

## IV. The Method

### *A. Motivation and two guiding principles*

Adaptive control of a dynamic network requires recomput ing the MDS for current temporal subgraph as the network structures evolve, but the new MDS may differ significantly from the previous MDS. Importantly, a temporal subgraph usu ally contains multiple MDSs [51], [52], each having different driver nodes (Figure 2). Therefore, we can select similar MDSs across adjacent temporal subgraphs to minimize ECC. For instance, Figure 3A shows a simple dynamic network and its sequential temporal subgraphs. In Figure 3B, the initial MDS of subgraph $G_1$ includes nodes $v_1$ and $v_2$. As the network evolves to $G_2$, the MDS shifts to nodes $v_3$ and $v_4$, indicating a full replacement of driver nodes. Over time, the network requires control over eight driver nodes, with a total of nine changes in the MDS. In contrast, Figure 3C represents a more stable MDS sequence with only four driver nodes and merely two changes of driver nodes between consecutive subgraphs. The presence of multiple selectable MDSs within a subgraph opens an avenue to optimizing the overall MDS of a dynamic network.

An effective approach to selecting the MDS for the current temporal subgraph must consider historical network topologies and possible future network changes. Historical topologies can be explored and exploited to estimate future network structures and help reduce ECC in two aspects. First, historically stable nodes, which often appeared in the node sets of the previous temporal subgraphs, are preferred to be included in the current MDS; doing so can minimize swapping in and out driver nodes and stabilize the overall adaptive control of the network. We refer to this consideration as *the stability principle of adaptive control*. Second, when computing the current MDS, we can explore the previous MDS to achieve the maximum consistency between the current and the previous MDSs, thereby minimizing the change to the MDS. We refer to this consideration as *the consistency principle of adaptive control*.

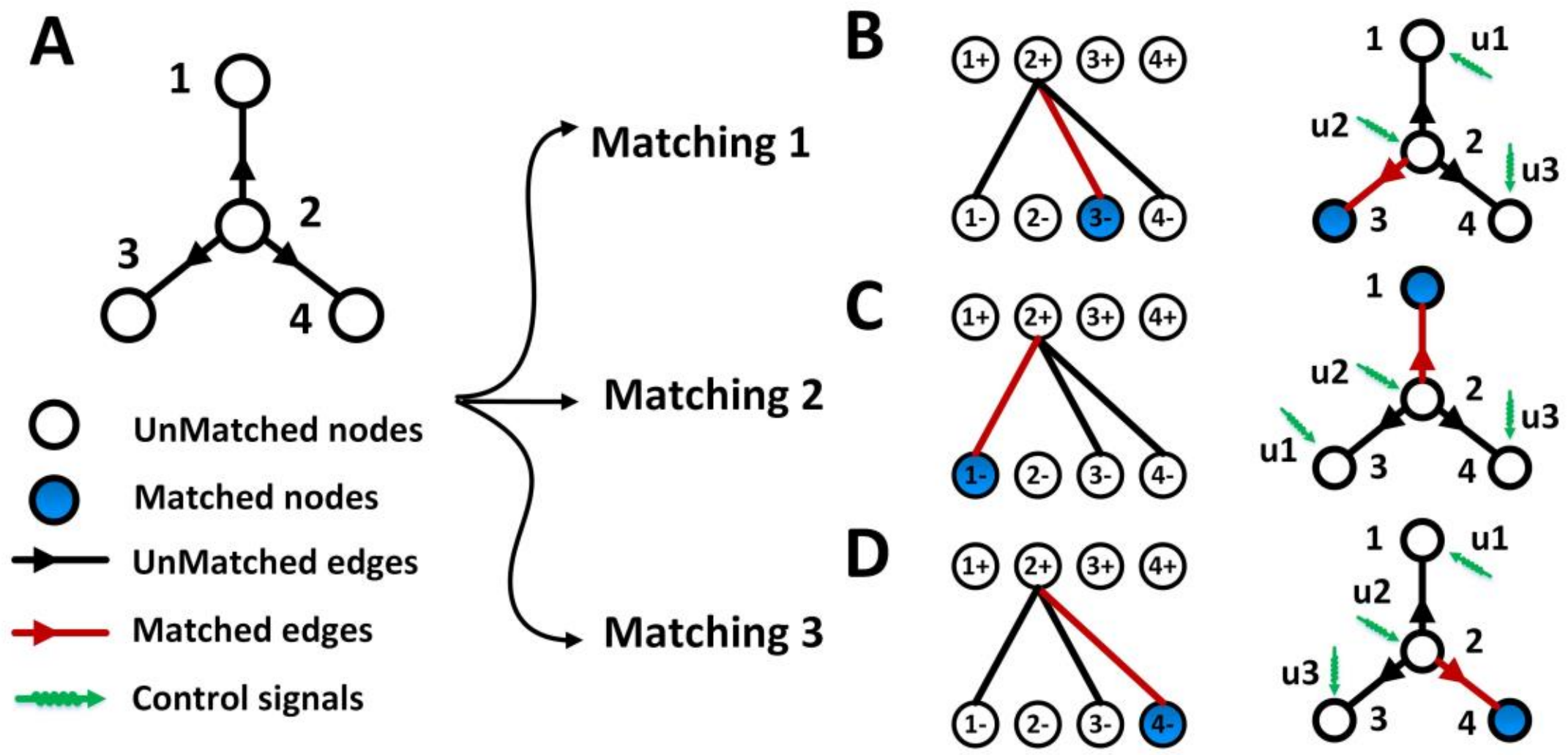

Fig. 2. A simple network with three different MDS. (A) A simple network with four nodes and three edges. (B)-(D) Different maximum matching of the network (A). Different maximum matchings may produce different MDSs.

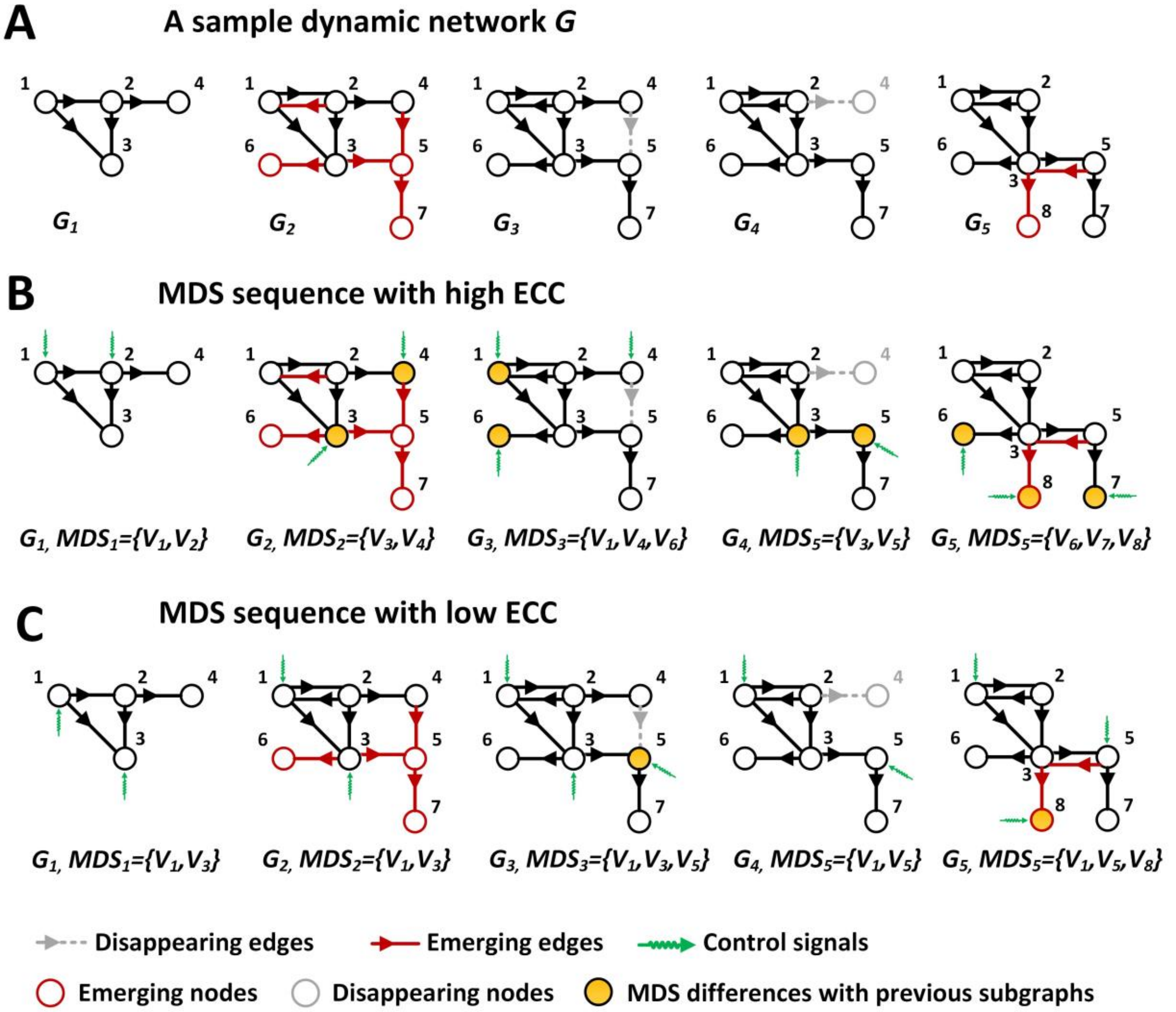

Fig. 3. Different MDS sequences of a network. (A) A sample dynamic network with five temporal subgraphs. (B) An MDS sequence with a high ECC, where eight driver nodes are required to control all five temporal subgraphs and nine driver node swappings between neighboring subgraphs are needed. (C) An MDS sequence with a low ECC, where four driver nodes and two driver node swappings are sufficient to control the five temporal subgraphs.

*B. Adaptive control metric of nodes*

We introduce an adaptive control metric to materialize these two principles to develop an MDS algorithm to minimize the variations to the resulting MDS and reduce the objective ECC. The adaptive control metric is designed to measure the stability and consistency of nodes in the current MDS with respect to historical network topologies. Following the stability principle, we measure the stability of node $v$ by considering two factors: 1) the number of edges that node v has in the current and all previous temporal subgraphs, i.e., the degree of node $v$; 2) the similarity of edges that node v has in adjacent temporal subgraphs. A higher node degree means that node v interacts with more nodes in the network, suggesting it is more stable [53], [54]. On the other hand, regardless of node degree, a node is also stable if its edges never change [55]. Therefore, we compute the similarity of edges that node v has in neighboring temporal subgraphs, i.e., $S_t(v) = \frac{|E_t(v) \cap E_{t-1}(v)|}{|E_t(v) \cup E_{t-1}(v)|}$, where $E_t(v)$ denotes the edge set of node v in temporal subgraph $G_t$. To sum up, the stability $\sigma_t(v)$ of node v in the current temporal subgraph $G_t$ can be computed as follows:

$$\sigma_t(v) = S_t(v) * DC_t(v), \qquad t > 1$$

where $S_t(v) \in [0,1]$ is the edge similarity of node $v$ in adjacent subgraphs. The degree centrality of node $v$ at time $t$ is given by $DC_t(v) = \frac{\sum_j x_{v,j}(t)}{N(t)-1} \in [0,2]$, where $x_{v,j}(t)$ is a binary variable that indicates whether a directed edge exists from node $v$ to node $j$ at time t. Specifically, $x_{v,j}(t) = 1$ if there is a directed edge from $v$ to $j$, and $x_{v,j}(t) = 0$ otherwise. $N(t)$ is the number of nodes in $G_t$. Since $G_t$ is a directed graph, if both $v$ and $j$ are mutually connected by bidirectional edges, they are counted twice, thus the range of degree centrality is [0,2].

Furthermore, as the network evolves continuously, we consider the stability of node $v$ over a period. We take the total stability of node $v$ across $l$ immediately previous temporal subgraphs as the final stability measure. The final stability $stb_t(v)$ of node $v$ in the current temporal subgraph $G_t$ can be computed as:

$$stb_t(v) = \sum_{i=t-l}^{t} \sigma_t(v) \qquad (1)$$

This stability principle rests on a fundamental assumption: nodes with higher degrees or more persistent connections tend to occupy more stable roles in real-world systems. Such nodes are less likely to disappear abruptly or fluctuate frequently across time steps. By prioritizing historically stable nodes, this strategy helps preserve the continuity of the driver node set and may enhance the robustness of the control structure under moderate topological variations.

To increase consistency between the MDSs of adjacent subgraphs (by the consistency principle), we prefer to keep the driver nodes in the last subgraph as the driver nodes of the current subgraph. Thus, the node consistency metric qv of $G_t$ is given as:

$$q_v = 2l * P_{t-1}(v) + stb_t(v) \qquad (2)$$

where $P_{t-1}(v)$ indicates whether node $v$ appears in the previous MDS, which is 1 if $v$ appears in the previous MDS or 0 otherwise:

$$P_{t-1}(v) = \begin{cases} 1, & v \in MDS_{t-1} \\ 0, & v \notin MDS_{t-1} \end{cases} \qquad (3)$$

We give the consistency principle greater importance than the stability principle, i.e., a node that appeared in the last MDS is prioritized as a driver node over nodes with high stability. The node consistency metric $q_v$ has the following property:

**Property 1:** Given two nodes $v_1$ and $v_2$ in a temporal subgraph $G_t$, if $v_1 \in MDS_{t-1}$ and $v_2 \notin MDS_{t-1}$, then $q_{v_1} > q_{v_2}$.

**Proof:** In $G_t$, since $S_t(v) \in [0,1]$ and $DC_t(v) \in [0,2]$, then $stb_t(v) \in [0,2l]$. From equations (2) and (3), node $v_1$ appears in $MDS_{t-1}$, then $q_{v_1} \in [2l,4l]$. On the other hand, since node $v_2$ does not appear in $MDS_{t-1}$, then $q_{v_2} \in [0,2l]$. Thus, $\max\{q_{v_2}\} \leq \min\{stb_{v1}\}$. Now, we prove that $\max\{q_{v2}\} = \min\{stb_{v1}\}$ is impossible. If this case is true, $q_{v1} = q_{v2} = 2l$, leading to $stb_t(v_1) = 0$ and $stb_t(v_2) = 2l$. $stb_t(v_1) = 0$ implies that $\sigma t(v1) = 0$ always holds in previous l temporal subgraphs, i.e., at least one of $DC_t(v_1) = 0$ or $S_t(v_1) = 0$ holds in each of the previous $l$ temporal subgraphs (denoted as condition A). $stb_t(v_2) = 2l$ implies that $DC_t(v_2) = 2$ is always true in previous $l$ temporal subgraphs (denoted as condition B). In other words, $\max\{q_{v2}\} = \min\{stb_{v1}\}$ implies that conditions A and B hold concurrently. However, condition B requires that node $v_2$ must always have the same connection to node $v_1$. In contrast, condition A means that node $v_2$ can never be connected to node $v_1$. Therefore, conditions A and B are mutually exclusive, i.e., $\max\{q_{v2}\} = \min\{stb_{v1}\}$ cannot be true. Therefore, $\max\{q_{v2}\} < \min\{stb_{v1}\}$.

*C. Adaptive control algorithm*

We develop the AC algorithm to minimize the total ECC of controlling a dynamic network. The algorithm is tailored to ensure consistency between the MDSs of sequential temporal subgraphs. A pivotal aspect of our approach is the integration of the adaptive control metric for nodes. This metric is used to prioritize nodes in computing maximum matching, ensuring an efficient and systematic approach to managing network dynamics.

In a nutshell, the AC algorithm leverages the incumbent maximum matching $M'$ from the previous graph $G'$ to effi ciently derive the maximum matching $M$ for the updated graph $G$. This algorithmic strategy is particularly beneficial when the graph undergoes minor and incremental changes, thereby minimizing disruption to its overall controlling structure. The algorithm focuses on the difference between $G'$ and $G$ by finding edges in $M'$ that are missing in $G$ and vice versa. Starting from the existing partial matching $M'$, it then sys tematically searches for augmenting paths to convert $G'$ to $G$ using the differential edges between G′ and $G$. This procedure is iteratively executed until the maximum matching for $G$ is determined.

Throughout this maximum matching process, the adaptive control metric is instrumental in ordering the nodes to be explored. This metric provides a well-defined strategy for node prioritization. The Breadth-First Search (BFS) algorithm, adopted in this phase, employs this node ordering to explore paths that engage more stable nodes. This focused approach not only addresses specific node preferences but also diligently maintains the coherence and consistency of the matching framework, adapting to the dynamic nature of the graph. In the worst case, where the adaptive control metric becomes ineffective due to abrupt structural changes, the AC algorithm reduces to a random order matching, which is equivalent to the normal maximum matching. In such cases, the ECC produced by the AC algorithm will be very close to that of the normal maximum matching, as both approaches rely on same principles when the adaptive control metric becomes ineffective. The pseudocode of the algorithm is listed in Algorithm 1.

Figure 4 illustrates the AC algorithm on an example dynamic network. Initially, in temporal subgraph $G_1$, the metric $\{q_1 = 0, q_2 = 0, q_3 = 0, q_4 = 0\}$ is calculated by formula (2), resulting in an MDS of $\{v_3,v_4\}$. Subsequently, when the network evolves to $G_2$, the node metric q is recalculated by formula (2), and an augmenting path search is performed by prioritizing the metric q in ascending order, leading to an MDS of $\{v_3,v_4\}$ (Figure 4A). This procedure is repeated until the network no longer undergoes any further changes.

We illustrate the new maximum matching using the tempo ral subgraph $G_2$ (Figure 4B). Initially, the matching edges in the previous temporal subgraph $G_1$ will be set as the initial matching edges of $G_2$ in the bipartite graph $b_0$ (i.e., edge $v_1 \rightarrow v_2$ and edge $v_2 \rightarrow v_1$), and the node search order is set to $\{v_7,v_6,v_5,v_1,v_2,v_4,v_3\}$. Next, the augmenting path of $v_7$ is searched for, resulting in $b_1$. This is followed by searching for the augmenting path of v6, reaching $b_2$. Subsequently, searching for an augmenting path of $v_5$ ends in $b_3$. Note that although $v_1$ and $v_2$ have already been matched at the beginning, once there is an augmented path connected to them in the process of $b_1$-$b_3$, $v_1$ and $v_2$ will no longer be matched. As a result, the unmatched nodes $\{v_3,v_4\}$ are determined to be the MDS of the temporal subgraph $G_2$.

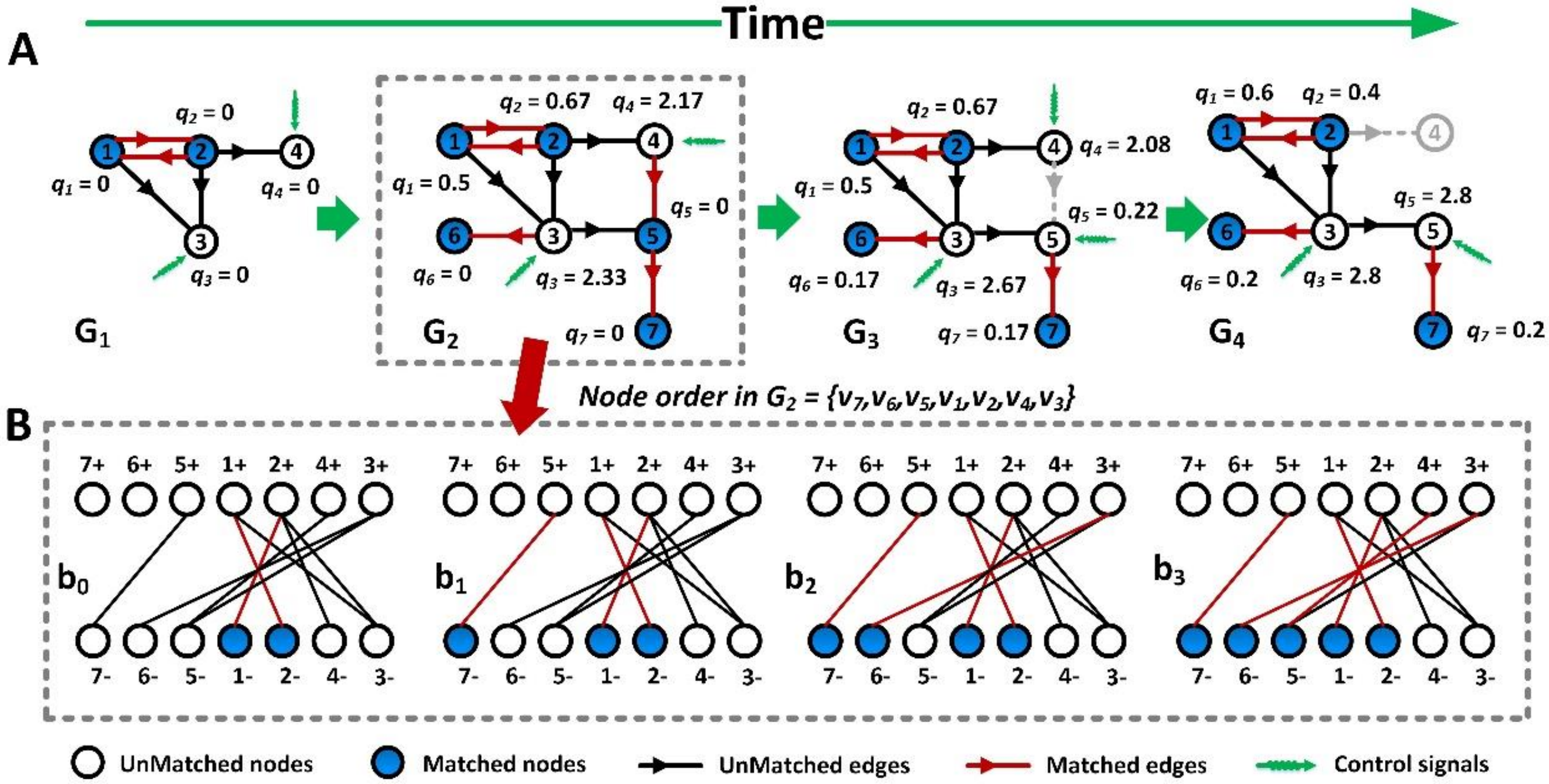

Fig. 4. An AC algorithm instance. (A) A simple dynamic network with four temporal subgraphs. (B) The process of maximum matching repair method of G2. In this instance, the parameter l of the adaptive control metric is 1.

**ALGORITHM 1: Adaptive Control Algorithm**

1. **Input**: current temporal subgraph *G*; previous temporal subgraph *G'*; previous matching *M'*; previous MDS *preMDS*; parameter *l*;
2. **Output**: current MDS *MDS*; current matching *M*;

// Step1 Remove Edges From Matching

3. *RemovedEdges* = getRemovedEdges(*G*, *G'*)

// Remove edges from *M'* that are no longer in *G*

4. **for** edge in *RemovedEdges* **do**:
5. **if** edge in *M'* **then**:
6. removeFromMatching(*M'*, edge)
7. **end if**
8. **end for**

// Step2 CalculateAdaptiveControlMetric

9. **for** each node *v* **in** *G* **do**:
10. stability[v] = degreeCentrality[v] * edgeSimilarity[v]
// stability, degreeCentrality and edgeSimilarity are all dictionary

11. **if** *v* in *preMDS* **then**
12. *P*[*v*]= 1
13. **else**
14. *P*[*v*]= 0
15. **end if**
16. q[*v*] = 2 * *l* * P[*v*] + stability[*v*]
17. **end for**

// Step3 GetNewMDS

18. *MDS*, *M* = GetMDS (*G*, *M', q*)

19. **Function** GetMDS(*G*, *M', q*)
20. **Get** order *q'* **by** ascending weight vector *q*;
21. **Repeat**
22. **Get** unmatched nodes set *un* **by** matching *M'*;
23. **For** node *n* **in** *un* **by** order *q'*:
24. **If** node *n* has the augmenting paths then:
25. Expand the augmenting paths and obtain a new matching *M**;
26. **end if**
27. **end for**
28. **Find** all unmatched nodes *un'* after matching;
29. **Let** *M* = *M**;
30. **Let** MDS = *un'*;
31. **Until** no augmenting path is found;
32. **Return** *MDS*; *M*;

## V. Experimental Results and Analysis

### *A. Network datasets and Experimental Environment*

We comprehensively evaluated the AC algorithm on syn thetic and real-world networks. A total of 1,860 synthetic networks were constructed based on two well-known models: the Scale-Free (SF) model [56] and the Erd˝os-R´enyi (ER) model [57]. We first utilized the SF model to generate 31 static network instances. Each instance contains 10,000 nodes. The average degree of the networks varied from 2.0 to 8.0 with an increment of 0.2 while holding the degree exponent constant at 3. Subsequently, 100 temporal subgraphs were created for each of these 31 static instances by partially reconnecting edges randomly with the degree exponent set to 3. The reconnection was determined by 30 reconnection ratios r, which varied between 0.01 and 0.30 in steps of 0.01. As a result, a total of 31 × 30 = 930 dynamic network instances were generated for the SF model, each consisting of 100 temporal subgraphs.

Similarly, for the ER model, we followed the identical procedure as described for the SF model, except reconnection was completely random, resulting in an additional 930 network instances. In summary, our experiments encompassed a vast synthetic dataset originating from the SF and ER models, providing a robust foundation for evaluating the efficacy of our algorithm.

We also analyzed AC on 20 real-world dynamic networks [58], [59], [60], including email, human contact, and social networks, which were sourced from previous studies (TABLE II). These networks exhibit diverse topological and temporal properties, making them particularly suitable for evaluating dynamic network control algorithms. Their diverse characteristics, spanning from smaller, focused group interactions to larger, complex structures, accommodate thorough benchmark ing across different scales and scenarios. These testbeds not only validate our new algorithm but showcase it in many real world applications.

All experiments were conducted on a server equipped with dual Intel Xeon Gold 5218R processors with 128 GB RAM, running Ubuntu 20.04 LTS and the algorithms were coded in Python.

**Table 2**. Real dynamic networks dataset. $N_{mean}$ is the mean number of nodes in the real dynamic network; $N_{min}$ and $N_{max}$ are the minimum and maximum numbers of nodes in the network, respectively. $k_{mean}$ is the mean value of the average degree of all temporal subgraphs of the dynamic network. $k_{min}$ and $k_{max}$ are the minimum and maximum average degrees of the network, respectively. $T$ is the number of temporal subgraphs contained in the dynamic network.

| *Type* | *Dynamic Networks* | $N_{mean}$ | $N_{min}$ | $N_{max}$ | $k_{mean}$ | $k_{min}$ | $k_{max}$ | *T* |
|---|---|---|---|---|---|---|---|---|
| *Email* | *email-Eu-core-temporal* | 763.22 | 738 | 818 | 19.62 | 17.79 | 21.93 | 9 |
| | *email-Eu-core-temporal-Dept1* | 231.67 | 219 | 239 | 9.94 | 8.81 | 11.99 | 9 |
| | *email-Eu-core-temporal-Dept2* | 120.78 | 115 | 127 | 13.28 | 11.57 | 14.55 | 9 |
| | *email-Eu-core-temporal-Dept3* | 72.67 | 66 | 81 | 11.94 | 10.32 | 13.56 | 9 |
| | *email-Eu-core-temporal-Dept4* | 108.11 | 102 | 120 | 11.36 | 9.44 | 12.63 | 9 |
| | *ia-radoslaw-email* | 93.94 | 2 | 140 | 7.64 | 1.0 | 14.89 | 118 |
| *Social* | *sx-mathoverflow* | 2038.02 | 812 | 2461 | 7.93 | 3.46 | 14.77 | 40 |
| | *sx-mathoverflow-a2q* | 1532.75 | 403 | 1866 | 3.36 | 1.48 | 8.49 | 40 |
| | *sx-mathoverflow-c2a* | 1185.18 | 356 | 1433 | 4.71 | 2.10 | 9.26 | 40 |
| | *sx-mathoverflow-c2q* | 1285.4 | 384 | 1628 | 5.01 | 2.90 | 7.45 | 40 |
| | *sx-superuser* | 5642.14 | 39 | 10246 | 4.07 | 1.6 | 10.24 | 93 |
| | *sx-superuser-a2q* | 4032.52 | 19 | 6429 | 2.33 | 1.26 | 6.23 | 93 |
| | *sx-superuser-c2q* | 2171.78 | 2 | 4352 | 3.04 | 1.0 | 3.71 | 93 |
| | *fb-forum* | 434.5 | 109 | 728 | 4.86 | 2.11 | 12.59 | 10 |
| | *CollegeMsg* | 631.86 | 200 | 1371 | 7.35 | 2.37 | 15.012 | 7 |
| *Human contact* | *contacts-prox-high-school-2013* | 291.33 | 232 | 312 | 12.90 | 5.68 | 16.6 | 6 |
| | *edit-enwikibooks* | 3453.56 | 2 | 14451 | 1.97 | 1.0 | 2.48 | 139 |
| | *copresence-LH10* | 31.54 | 10 | 48 | 15.32 | 3.4 | 26.65 | 13 |
| | *ia-workplace-contacts* | 73.75 | 62 | 87 | 6.82 | 3.03 | 10.94 | 4 |
| | *ia-hospital-ward-proximity-attr* | 42 | 32 | 49 | 12.56 | 6.94 | 17.49 | 8 |

### *B. Performance metric and benchmark algorithm for comparison*

The temporal subgraphs of a real network may follow vari able rates of change. These fluctuations may significantly af fect algorithm performance. To address this issue, we adopted the Jaccard index to quantify the overlap between nodes or edges in consecutive temporal subgraphs, thereby measuring their temporal consistency. In particular, we employed the Jaccard similarity coefficient [61] to quantify the node sim ilarity and structural similarity between neighboring temporal subgraphs, expressed as $J(A, B) = \frac{|A \cap B|}{|A \cup B|}$ where A and B are the node sets (or edge sets) of subgraphs $G_{i-1}$ and $G_i$, respectively.

To establish a sound benchmark to evaluate our AC strat egy and algorithm, we revised and enhanced an algorithm originally designed for static networks proposed by Liu et al. [18]. Specifically, we adopted the maximum matching based algorithm (MM), which has been demonstrated to be state-of-the-art in computing the MDS for static networks. The MM algorithm is applied independently to each temporal subgraph to generate an MDS sequence, serving as a baseline for comparison with the AC algorithm. Using the MM algo rithm as the comparison benchmark has several advantages. Firstly, adaptive control is a completely new problem, and there is still no suitable algorithm to solve it. Secondly, MM computes exact solutions for the MDS problem, which is critically important for dynamic network control. In contrast, many existing dynamic network control algorithms compute approximate MDS solutions, often resulting in suboptimal control performance [33], [34], [36], [35]. Furthermore, the revised MM algorithm does not rely on knowledge of network changes over time. This characteristic aligns well with the scenario we consider where knowledge of network alterations is typically unavailable. In contrast, many dynamic control algorithms require a comprehensive understanding of the entire network's temporal dynamics, which can be a significant limitation in practice.

*C. Results on synthetic networks*

The AC and MM algorithms were first experimentally compared on synthetic dynamic networks (Figure 5). The ECC for the AC algorithm, $ECC_{AC}$, the MM algorithm, $ECC_{MM}$, and their ratio $ECC_{AC}/ECC_{MM}$ were computed. AC consistently outperformed MM on synthetic ER and SF networks with varying reconnection ratio r and average degree k, achieving $ECC_{AC}/ECC_{MM}<1$ on the two types of synthetic networks we tested (Figure 5). A smaller $ECC_{AC}/ECC_{MM}$ ratio indicates that AC outperforms MM. Notably, on stable networks (with reconnection ratio $r < 0.1$), the AC was nearly twice as effective as the MM algorithm, spending almost half of the ECC. Even with an increasing reconnection ratio r, the AC algorithm outperformed MM across networks with varied average degrees. Furthermore, similar optimization outcomes across various parameter $l$ values indicated a specific trend in MDS optimization selection. It appeared that the MDS optimization selection might depend on network topological structures and that of the immediately precedingtemporal subgraph.Consequently, there isnoneed to retain the full history of topology, which enhanced the efficiencyof thealgorithmandbroadeneditsapplicability.

Moreover, the AC algorithm was more efficient on SF-based dynamic networks, averaging a ratio of $ECC_{AC}/ECC_{MM}$ =0.72, than on ER-based dynamic networks, with an average $ECC_{AC}/ECC_{MM}$ =0.84. This variance can be primarily attributed to the unique characteristics inherent to each network model. As per Equation (2), the AC algorithm preferentially favors nodes with greater stability. It aligns well with the power-law distribution trait in the SF model. In contrast, the ER model's random evolution cannot satisfy this characteristic, particularly when the reconnection ratio $r$ is large. Moreover, while AC's performance deteriorates with increasing reconnection ratio $r$, the decrement is more subtle for SF-based networks than ER-based networks.

We conducted an in-depth analysis at the temporal subgraph level to investigate the efficacy of our algorithm further. We compared the ECCs of the AC and MM algorithms on temporal subgraphs (Figure 6). The results on the ER-based dynamic networks showed that AC outperformed MM for 95% of the temporal subgraphs when the average degree $k$ was less than 4 (Figure 6A). However, as the average degree $k$ surpassed 4, temporal subgraphs displaying equivalent performance between the two algorithms increased (yellow part in Figure 6A). The results on the SF-based dynamic networks were compatible with those on the ER networks. The AC algorithm was superior to MM in 100% of the temporal subgraphs with an average degree $k$ below 6. As the average degree increases, the two algorithms performed similarly on more temporal subgraphs (yellow part in Figure 6B).

Notably, as network density increased, the discernible advantage of the AC algorithm relative to MM diminished. This is not indicative of AC's inadequacy for dense networks. When networks became denser, the MDS sizes decreased significantly (Figure 7). This reduction led to a constrained optimization space for the MDS within each temporal subgraph. As a result, both MM and AC algorithms may produce similar solutions. For instance, the mean $ECC_{AC}/ECC_{MM}$ ratio of ER-based dynamic networks was 0.63 when $k$=2.0 and elevated to 0.85 when $k$=8.0. We illustrate two representative extreme cases of temporal subgraphs in Figure 8, where the performance of AC was either equivalent to or worse than MM. They respectively explained the situations when the two algorithms performed equally and when MM outperformed AC. In fact, on ER-based networks, the two algorithms performed almost the same once the network reconnection ratio exceeded 0.50.

In dynamic networks with a high change rate in neighboring temporal subgraphs, the performance of the AC algorithm does not deteriorate monotonically as the temporal change rate increases. Instead, it has a lower-bound performance equivalent to the MM algorithm. Because the latter searches for augmenting paths in a fixed or random order, which is unbiased. When the adaptive control metric in AC becomes ineffective due to the high change rate, AC will behave the same as MM and perform matching by the default order. In this case, even if there is a slight discrepancy in the ECC results of the two algorithms in a single temporal subgraph, the overall ECC of the dynamic network will be similar. These boundary analyses collectively demonstrate that the AC algorithm possesses a well-defined performance lower bound, highlighting the conservative nature of its design and the robustness of its overall control strategy. In summary, these findings demonstrate that while the performance of AC is sensitive to network variability, its design ensures stable behavior and graceful degradation under dynamic conditions.

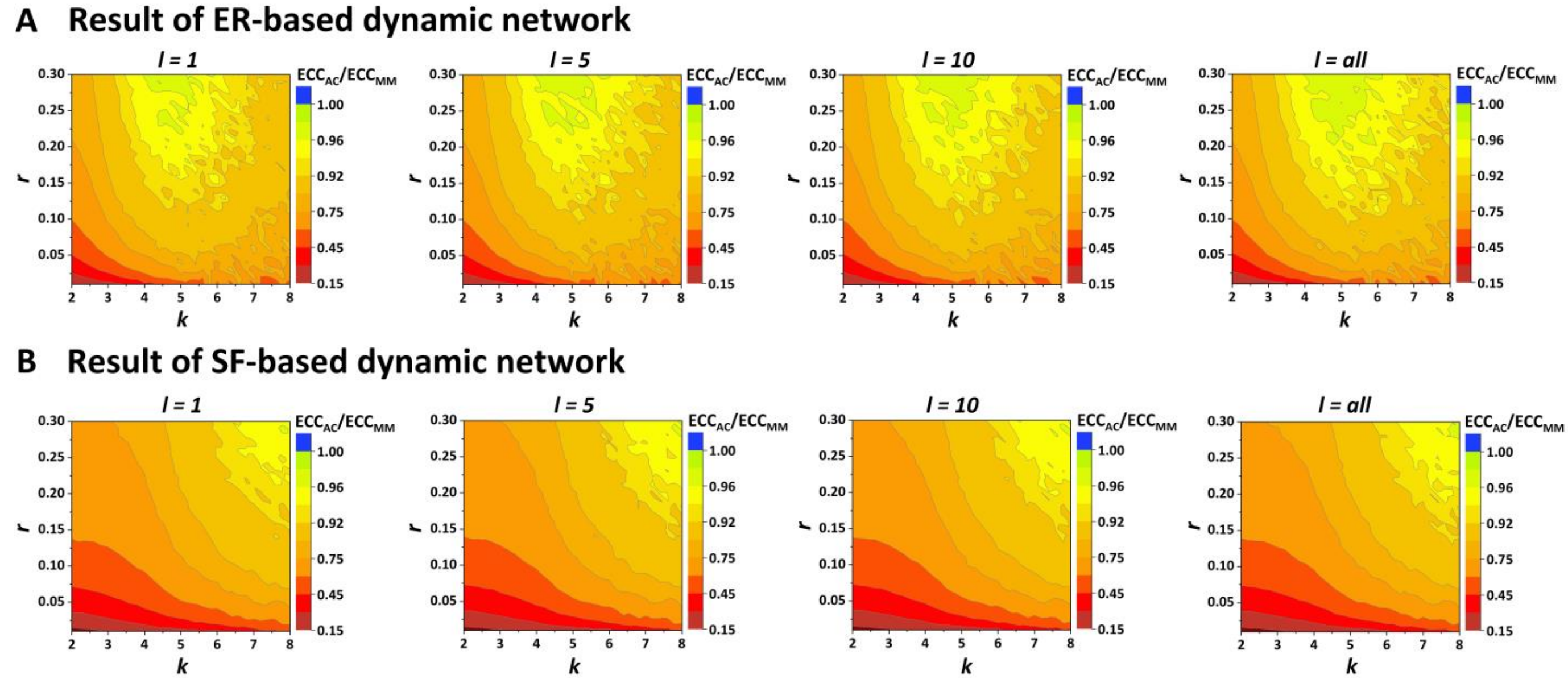

Fig. 5. The ECC results of the AC and MM algorithms on synthetic networks. (A) The result of ER-based dynamic network. (B) The result of SF-based dynamic network. The horizontal axis represents the average degree *k* of the network and the vertical axis represents the change rate *r*. The color scale indicates the value of the ratio $ECC_{AC}/ECC_{MM}$, with smaller values indicating superior algorithmic performance.

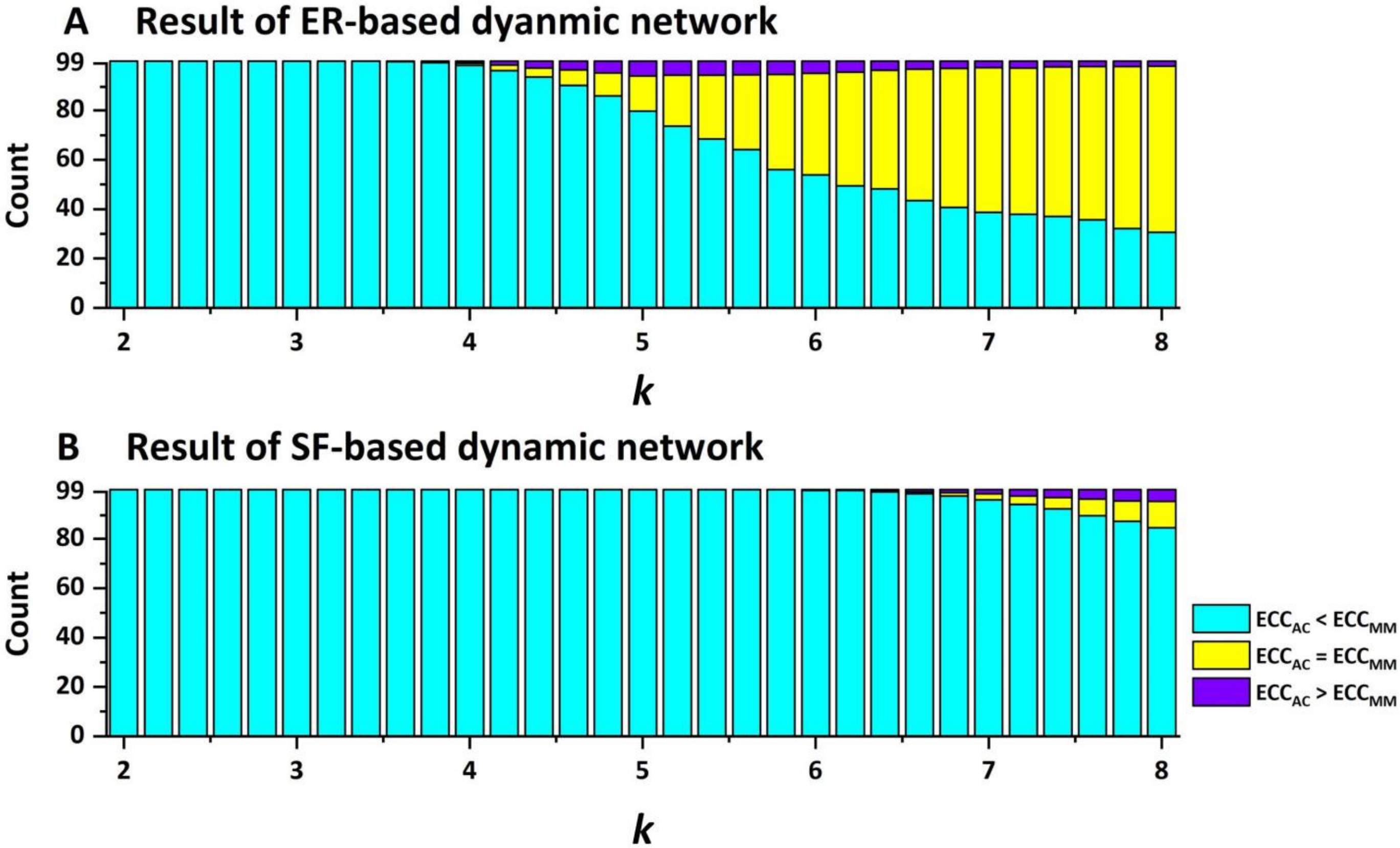

Fig. 6. Performance of the AC algorithm on all temporal subgraphs. (A) The performance of the AC on all temporal subgraphs in ER-based dynamic network. (B) The performance of the AC on all temporal subgraphs in SF based dynamic network. The horizontal axis represents the average degree k of the network and the vertical axis represents the average number of temporal subgraphs. For any given average degree *k*, the average temporal subgraphs is computed based on the average count of all temporal subgraphs with different change rates r and different number of pre-sequence temporal subgraphs considered *l*.

*D. Results on real networks*

We evaluated the AC and other algorithms on a diverse set of 20 real-world dynamic networks, chosen based on their topological structures and dynamic changes. The use of these networks was to ensure that the algorithms were evaluated across a wide range of conditions to gain a deep understanding of their generality and efficacy on real-world networks.

We computed the ECCs for the AC and MM algorithms on the 20 real networks. The results showed that AC outperformed MM, achieving an average $ECC_{AC}/ECC_{MM}$ ratio of 0.81 across all networks (TABLE III). However, the performance of AC varied significantly across different networks. For example, on the *email-Eu-core-temporal* and *contacts-prox-high-school-2013* networks, AC performed significantly better than MM, resulting in $ECC_{AC}/ECC_{MM}$=0.50. Conversely, AC and MM exhibited comparable performance on highly dynamically changing networks, like like *sx-superuser* and *edit-enwikibooks* (TABLE III).

To further understand the varying performance of the AC algorithm on different networks, we studied the dynamic network characteristics. We calculated the node and edge similarities of consecutive subgraphs as a representation of the network's dynamism (see Section Performance metric and benchmark algorithm for comparison). On networks exhibiting low dynamism, characterized by higher node and edge similarities, AC significantly outperformed MM. However, as the dynamism of the networks increased, the efficacy of AC declined. There is a negative correlation between the $ECC_{AC}/ECC_{MM}$ ratio and node and edge similarities (Figure 9). The higher the average similarity in consecutive subgraphs, the better that AC would perform. This phenomenon can be attributed to the fact that when there are drastic alterations in neighboring subgraphs, the fundamental objective of the AC optimization algorithm—aiming to preserve consistency in the driver node set across time—becomes challenging to achieve, resulting in a performance dip. However, in scenarios where the network undergoes gradual changes, AC distinctly surpasses MM, attesting to the efficacy of the AC algorithm.

Using the same analytic strategy as used for synthetic networks, we also analyzed the temporal subgraphs to understand AC's performance on real-world networks. Notably, AC exhibited superior performance over MM across all temporal subgraphs in 11 out of the 20 real networks analyzed (Figure 10). For 17 of these 20 networks, AC surpassed MM in over 80% of the temporal subgraphs. Figure 11 shows a more detailed view of the results for each temporal subgraph. Note that AC's performance fluctuates across different temporal subgraphs, even within the same real dynamic network (Figure 11).

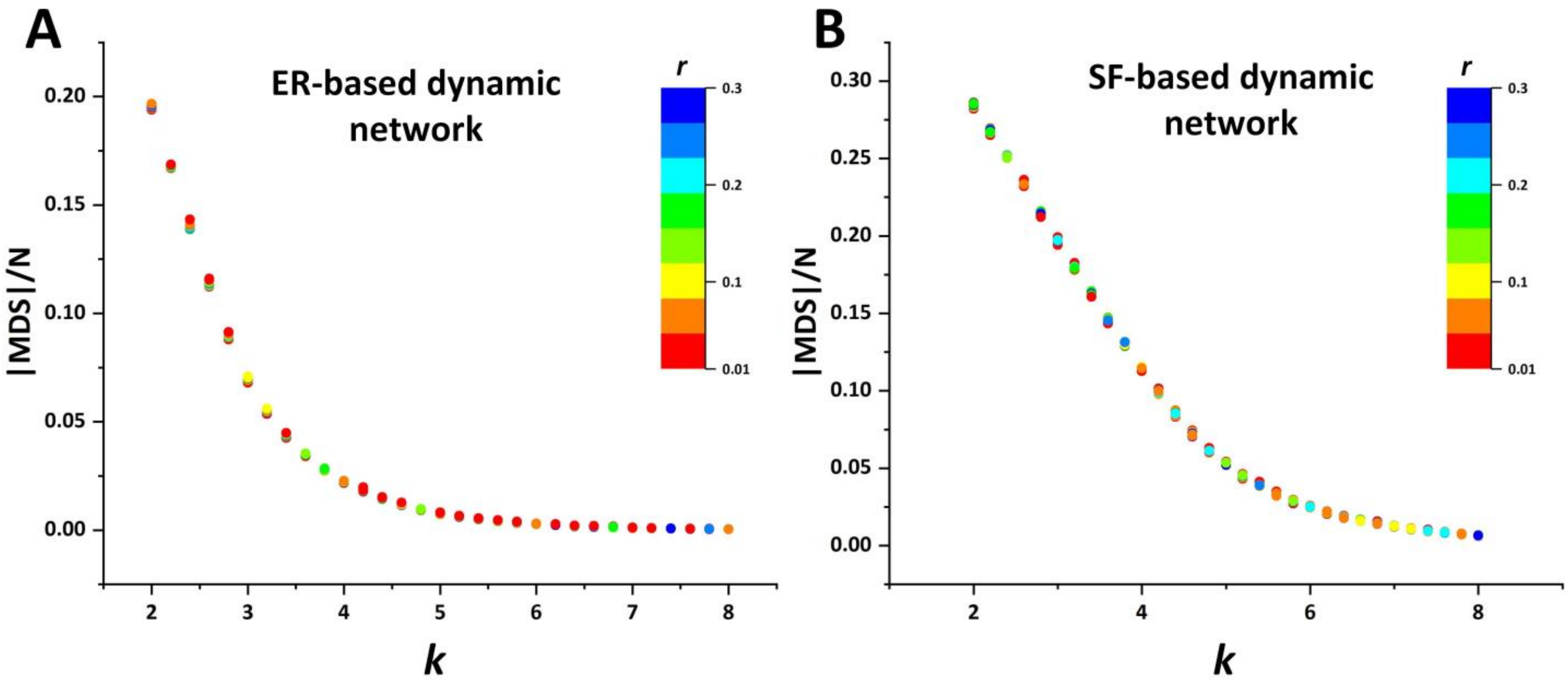

Fig. 7. The relationship between the MDS size and the average degree of the network. (A) The average MDS size in ER-based dynamic network. (B) The average MDS size in SF-based dynamic network. The horizontal axis shows the average degree k, while the vertical axis represents the average MDS ratio across all temporal subgraphs.

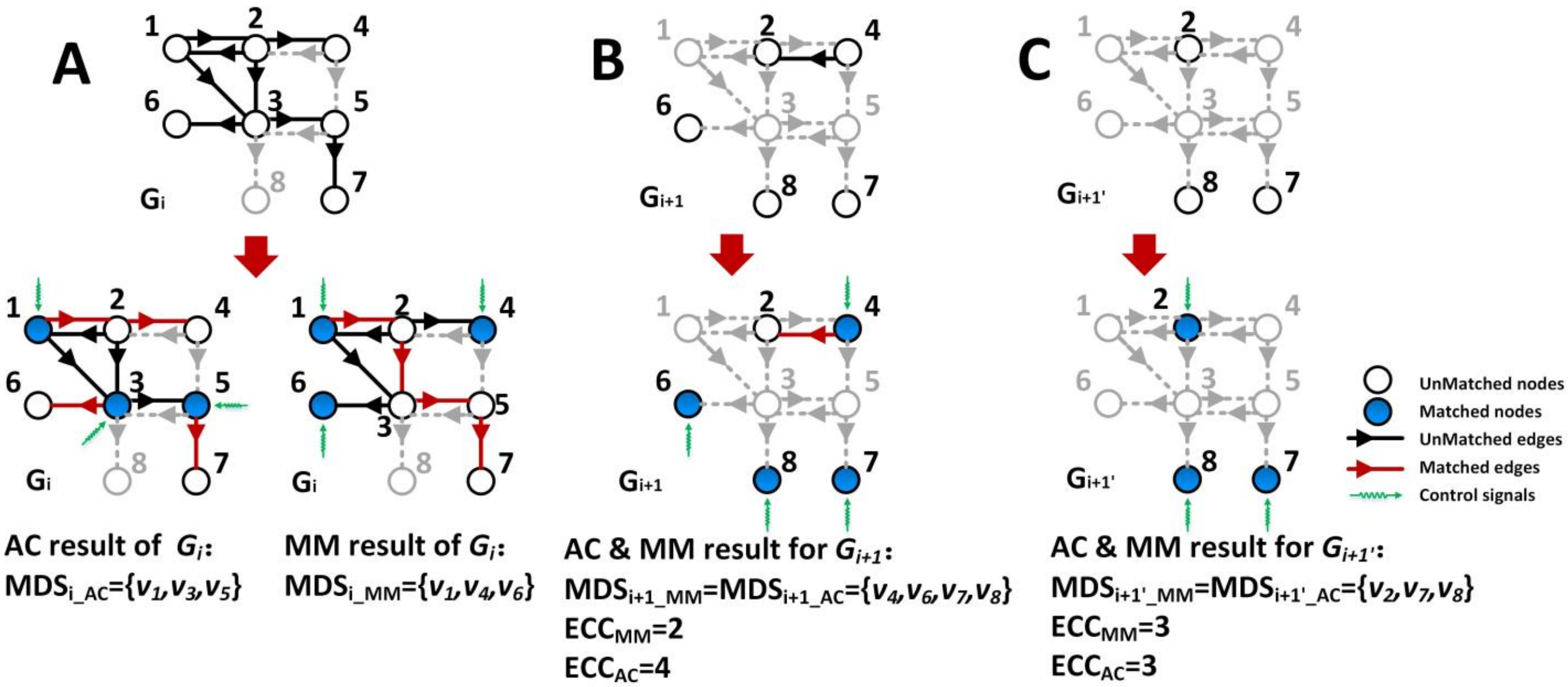

Fig. 8. Examples of extreme temporal subgraph variations. (A) A temporal subgraph Gi, with the MDS calculated by the AC and MM algorithms being $v_1,v_3,v_5$ and $v_1,v_4,v_6$, respectively. (B) Temporal subgraph $G_{i+1}$, where the MM algorithm outperforms the AC algorithm in terms of ECC optimiza tion efficiency. (C) Temporal subgraph $G_{i+1'}$, where the AC algorithm is equally efficient as the MM algorithm in terms of ECC optimization efficiency.

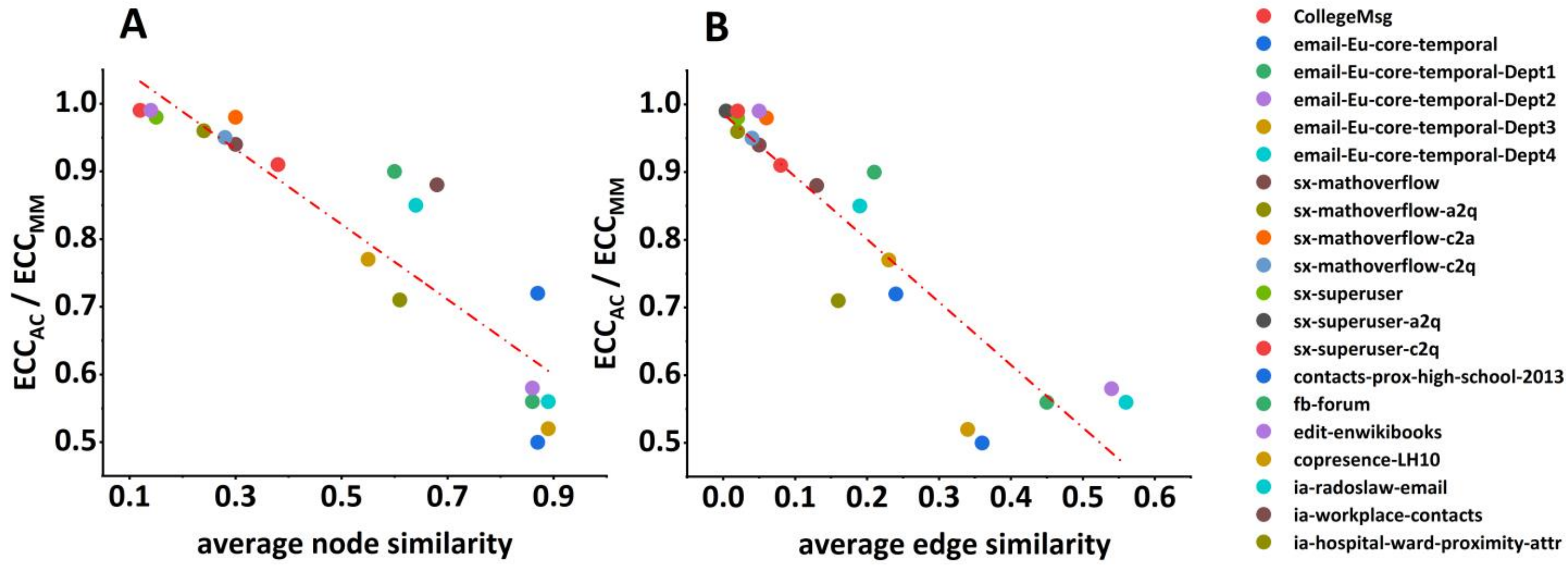

Fig. 9. Relationship between AC algorithm performance and similarity of temporal subgraphs in real networks. (A) The relationship between the average node similarity and AC algorithm performance. (B) The relationship between the average edge similarity and AC algorithm performance. The smaller $ECC_{AC}/ECC_{MM}$ means the higher algorithm performance.

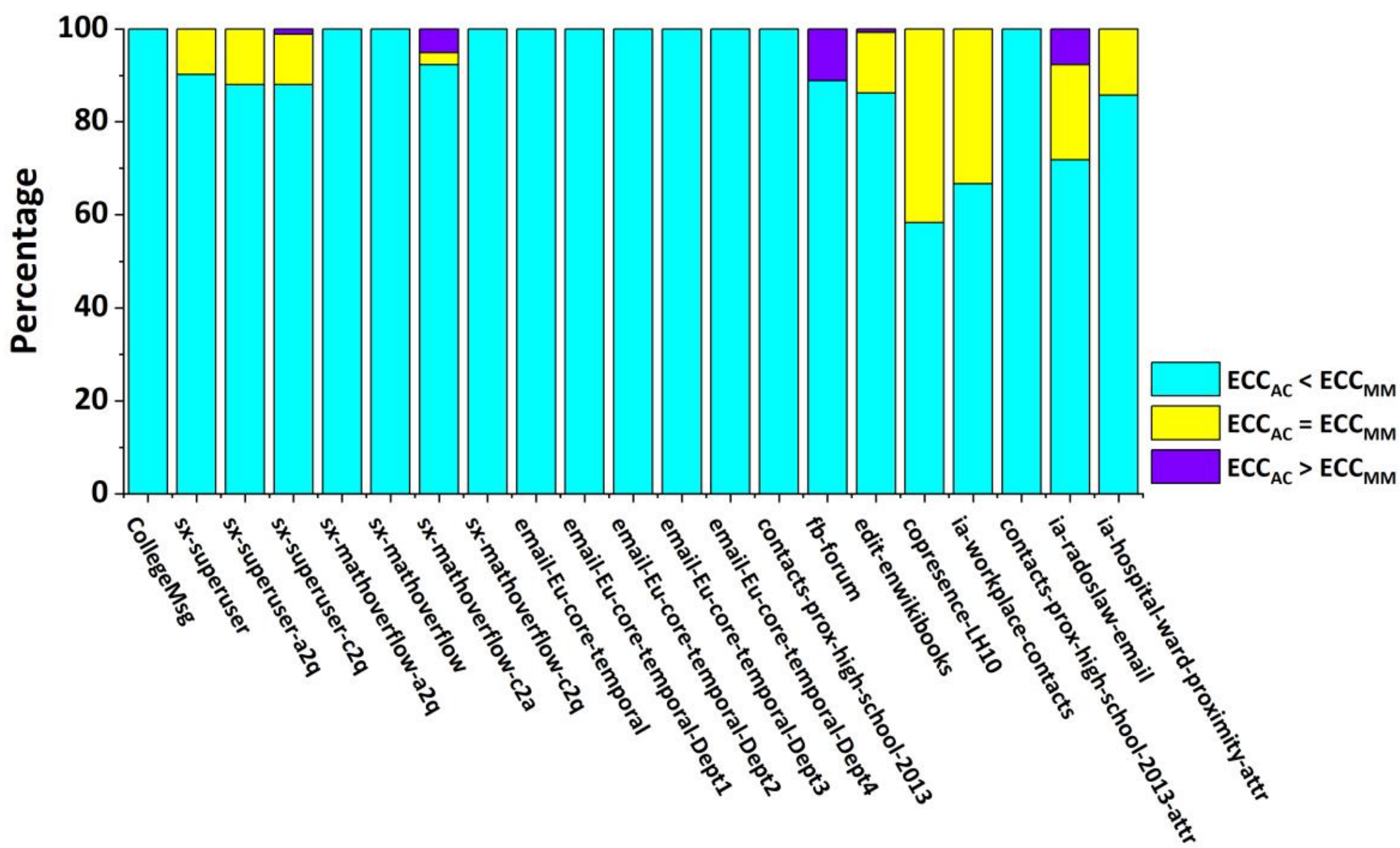

Fig. 10. Performance of the AC algorithm on temporal subgraph level in real network. The horizontal axis represents different real networks.

**Table 3.** Results on real networks. $S_n$ and $S_e$ denote the mean similarities of the node and edge of the networks, respectively.

| *Dynamic Networks* | $S_n$ | $S_e$ | *Average MDS* | $ECC_{AC}$ | $ECC_{MM}$ | $ECC_{AC}/ECC_{MM}$ |
|---|---|---|---|---|---|---|
| *CollegeMsg* | 0.38 | 0.08 | 236.71 | 934 | 1023 | 0.91 |
| *email-Eu-core-temporal* | 0.87 | 0.36 | 148.33 | 406 | 811 | 0.5 |
| *email-Eu-core-temporal-Dept1* | 0.86 | 0.45 | 55.78 | 154 | 275 | 0.56 |
| *email-Eu-core-temporal-Dept2* | 0.86 | 0.54 | 20.11 | 68 | 117 | 0.58 |
| *email-Eu-core-temporal-Dept3* | 0.89 | 0.34 | 13.78 | 43 | 82 | 0.52 |
| *email-Eu-core-temporal-Dept4* | 0.89 | 0.56 | 16.44 | 55 | 99 | 0.56 |
| *sx-mathoverflow* | 0.3 | 0.05 | 694 | 21508 | 22919 | 0.94 |
| *sx-mathoverflow-a2q* | 0.24 | 0.02 | 874.1 | 24279 | 25320 | 0.96 |
| *sx-mathoverflow-c2a* | 0.3 | 0.06 | 578.9 | 18246 | 18536 | 0.98 |
| *sx-mathoverflow-c2q* | 0.28 | 0.04 | 498 | 12018 | 12634 | 0.95 |
| *sx-superuser* | 0.15 | 0.02 | 2496.41 | 207738 | 211540 | 0.98 |
| *sx-superuser-a2q* | 0.12 | 0.004 | 2385.38 | 189038 | 191514 | 0.99 |
| *sx-superuser-c2q* | 0.12 | 0.02 | 927.6 | 67131 | 68091 | 0.99 |
| *contacts-prox-high-school-2013* | 0.87 | 0.24 | 52.5 | 96 | 134 | 0.72 |
| *fb-forum* | 0.6 | 0.21 | 212.6 | 815 | 908 | 0.9 |
| *edit-enwikibooks* | 0.14 | 0.05 | 2962.58 | 334285 | 337951 | 0.99 |
| *copresence-LH10* | 0.55 | 0.23 | 4.85 | 33 | 43 | 0.77 |
| *ia-radoslaw-email* | 0.64 | 0.19 | 22.22 | 1706 | 2017 | 0.85 |
| *ia-workplace-contacts* | 0.68 | 0.13 | 20 | 37 | 42 | 0.88 |
| *ia-hospital-ward-proximity-attr* | 0.61 | 0.16 | 15.5 | 45 | 63 | 0.71 |

### *E. Extended Comparative Analysis with Multiple Baseline Methods*

In the proposed AC algorithm, we introduce a preference parameter $q$ to guide the matching process, thereby generating more centralized and stable MDS compared to conventional approaches. To rigorously validate the effectiveness of this design, we benchmark AC against three representative control strategies: (1) the non-preferential maximum matching algorithm (MM), which is the basic baseline in previous results; (2) degree centrality preference-based algorithm (DPB), prioritizing high-degree nodes during maximum matching; (3) pagerank preference-based algorithm (PPB), prioritizing high-pagerank value nodes during maximum matching; This multi-perspective comparative framework enables systematic evaluation of AC's advantages in balancing optimality, adaptability, and computational efficiency across dynamic network scenarios.

The MM, DPB, and PPB algorithms employ distinct weight ing strategies for the parameter $q$ to guide subsequent maxi mum matching processes. In the MM algorithm, the weighting parameter $q$ is unassigned, leading to default maximum match ing without any node preference. For DPB, we utilize node degree [62] as the weighting criterion $q$, where nodes with smaller degrees are prioritized during matching. This weight ing strategy reserves high-degree nodes as candidates for the MDSin the current subgraph. In PPB, the PageRank centrality metric [63] serves as the weighting parameter $q$, ensuring that nodes with higher influence scores are systematically retained in the current MDS.

We adopt three key metrics for performance assessment: the size of the MDS (denoted as |MDS|) solved at each temporal subgraph; the union of MDS across all temporal sub graphs (UMDS); and the ECC, measuring cumulative driver node changes throughout the dynamic network evolution. Our experiments utilize ER random networks with 1,000 nodes, systematically configured with average degrees spanning 4.0 to 6.0 in 0.2 increments and topological variation rates spanning 0.10 to 0.20 in 0.1 increments. This parameter space gener ates 121 distinct initial network instances, each subsequently evolving through 100 sequential topological modifications to produce corresponding dynamic network realizations.

The comparative results of four algorithms are summarized in Table IV. For AC, MM, DPB, and PPB — all operating under the maximum matching framework — they generate identical |MDS| values since each precisely solves the MDS for every temporal subgraph. However, AC demonstrates superior performance in dynamic networks by achieving smaller UMDS (indicating more centralized driver nodes) and lower ECC (fewer control adjustments over time) across varying average degrees $k$ and topological variation rates $r$. The comparison among these algorithms is intended to highlight the effectiveness and robustness of the stability metric used in the AC algorithm, as their only difference lies in the design of node stability metrics. The results suggest that AC has a superior ability to identify structurally persistent and functionally important nodes in evolving networks.

**Table 4.** Performance Comparison of Different Algorithms (Mean ± Std)

| Parameter | Metric | Algorithm | | | |
|---|---|---|---|---|---|
| | | AC | MM | DPB | PPB |
| $k$ | \|MDS\| | 16.81±4.93 | 16.81±4.93 | 16.81±4.93 | 16.81±4.93 |
| [4.0−5.0) | UMDS | **518.77±99.82** | 534.92±95.77 | 539.73±96.82 | 554.13±96.29 |
| | ECC | **821.77±261.71** | 918.06±235.99 | 904.67±256.14 | 918.60±259.00 |
| $k$ | \|MDS\| | 5.84±1.86 | 5.84±1.86 | 5.84±1.86 | 5.84±1.86 |
| [5.0−6.0] | UMDS | **257.80±68.65** | 277.41±71.17 | 281.29±71.69 | 287.14±73.23 |
| | ECC | **327.82±100.97** | 367.39±113.13 | 367.69±112.56 | 365.44±111.54 |
| $r$ | \|MDS\| | 11.67±6.86 | 11.67±6.86 | 11.67±6.86 | 11.67±6.86 |
| [0.1−0.15) | UMDS | **356.86±146.35** | 376.27±148.79 | 379.68±147.74 | 377.04±146.04 |
| | ECC | **519.66±270.34** | 594.86±315.98 | 581.86±303.15 | 555.05±296.19 |
| $r$ | \|MDS\| | 10.74±6.36 | 10.74±6.36 | 10.74±6.36 | 10.74±6.36 |
| [0.15−0.2] | UMDS | **406.31±159.57** | 423.05±154.37 | 428.13±156.42 | 434.71±161.88 |
| | ECC | **605.44±325.59** | 666.00±354.76 | 664.98±349.36 | 668.39±358.45 |

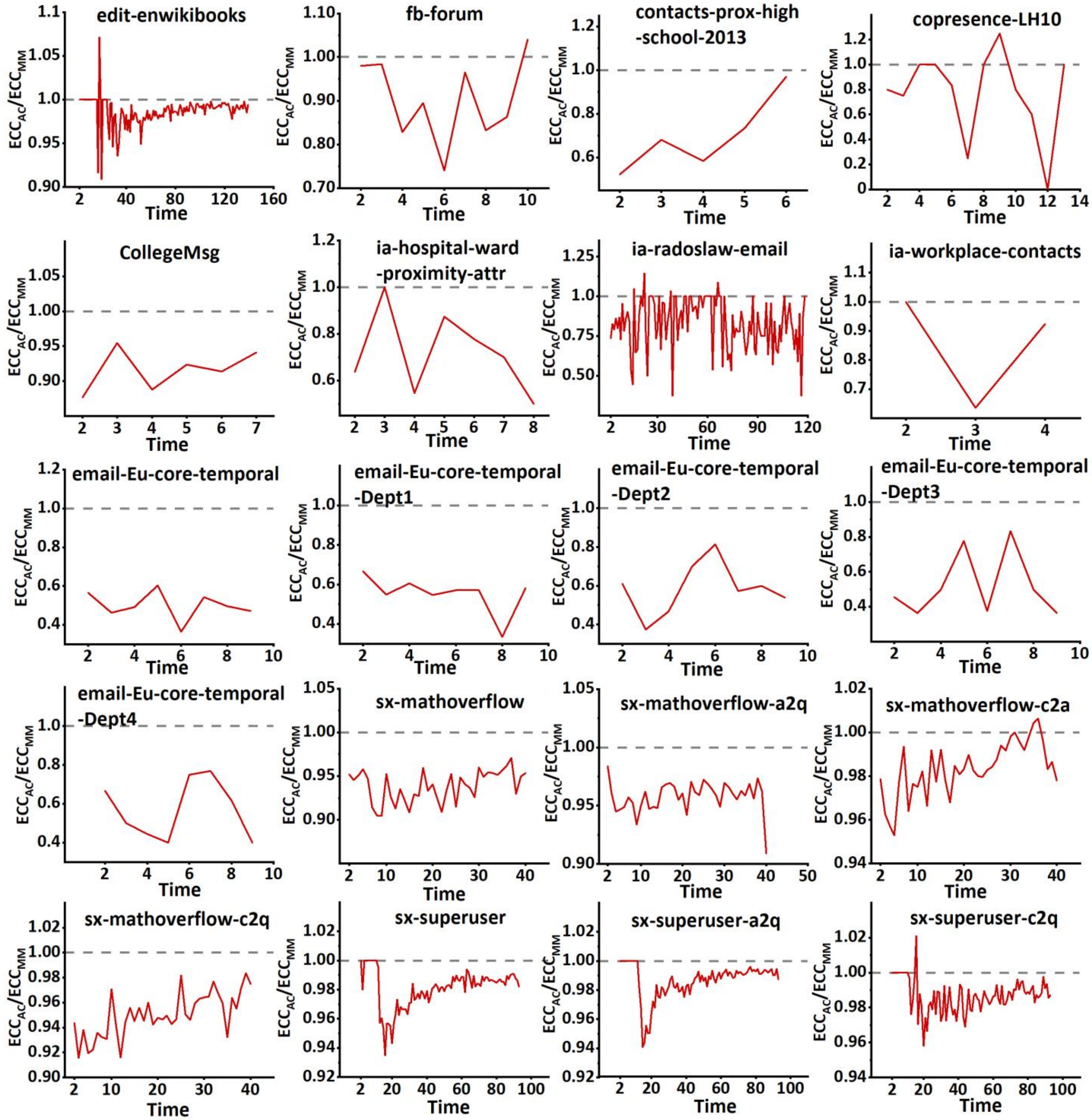

Fig 11. Optimization of the ECC for temporal subgraphs in real networks. The horizontal axis represents the temporal subgraphs at different time points, and the vertical axis represents the ECC ratios between the AC and MM algorithms. The gray dashed line with y=1 serves as the baseline, where the region above the line, i.e., $ECC_{AC}/ECC_{MM} > 1$, indicates the occurrence of negative optimization.

## VI. Conclusion

In this study, we formulated the adaptive control problem for dynamic networks and proposed a real-time control method, AC, to generate effective control schemes. We evaluated the proposed AC algorithm on both synthetic and real-world dynamic networks. A major result of our study is that the similarity of the consecutive temporal subgraphs that represent the evolving dynamics of a dynamic network has a great impact on how the network should be controlled and the performance of the AC algorithm. We further analyzed the performance bound of the AC algorithm and demonstrated that it outperforms the baseline MM algorithm on most tested dynamic networks. Importantly, AC maintains a well-defined performance lower bound—under worst-case conditions, it performs no worse than the MM algorithm. As the AC algorithm is heuristic in nature, we acknowledge that it currently lacks a formal approximation bound with respect to the global optimum of ECC. Future work could explore this direction by developing theoretical frameworks that quantify the performance gap between the proposed heuristic and the optimal adaptive control sequence, particularly under varying assumptions about network evolution. Overall, this study has not only provided an effective and practical method for adaptively controlling dynamic networks but also deepened our understanding of dynamic network control and opened new venues for network research and applications.

## AUTHOR CONTRIBUTION

XZ conceived the study and drafted manuscript. WZ revised the manuscript. CP, HZ, YZ and CZ implemented the algorithm, performed data analysis, prepared the figures and drafted the manuscript. ZS collated the data and revised the figures. All authors contributed to the preparation of the manuscript.

## ACKNOWLEDGMENT

This work was supported by the National Natural Science Foundation of China [62176129], the National Natural Science of China-Jiangsu Joint Fund [U24A20701], the Hong Kong RGC Strategic Target Grant [grant number STG1/M-501/23-N], the Hong Kong Health and Medical Research Fund [grant number 10211696], the Hong Kong Global STEM Professor Scheme, and the Hong Kong Jockey Club Charities Trust.

## REFERENCES

[1] M.E.J.Newman,"The Structure and Function o fComplex Networks," SIAM Review,vol.45,no.2,pp.167–256, Jan.2003.

[2] S. Boccaletti, V. Latora, Y. Moreno, M. Chavez, and D. U. Hwang, "Complex networks: Structure and dynamics," Physics Reports, vol. 424, no. 4, pp. 175–308, Feb. 2006.

[3] S. H. Strogatz, "Exploring complex networks," Nature, vol. 410, no. 6825, pp. 268–276, Mar. 2001.

[4] C. Pan, Y. Zhu, M. Yu, Y. Zhao, C. Zhang, X. Zhang, and Y. Yao, "Control Analysis of Protein-Protein Interaction Network Reveals Po tential Regulatory Targets for MYCN," Frontiers in Oncology, vol. 11, p. 633579, Apr. 2021.

[5] X. Zhang, "Altering Indispensable Proteins in Controlling Directed Human Protein Interaction Network," IEEE/ACM Transactions on Com putational Biology and Bioinformatics, vol. 15, no. 6, pp. 2074–2078, Nov. 2018.

[6] X. Zhang, C. Pan, X. Wei, M. Yu, S. Liu, J. An, J. Yang, B. Wei, W. Hao, Y. Yao, Y. Zhu, and W. Zhang, "Cancer-Keeper Genes as Therapeutic Targets," iScience, vol. 26, no. 8, p. 107296, Aug. 2023.

[7] X. Wei, C. Pan, X. Zhang, and W. Zhang, "Total network controllability analysis discovers explainable drugs for covid-19 treatment," Biology Direct, vol. 18, no. 1, p. 55, 2023.

[8] D. Zhou, Y. Kang, D. Cosme, M. Jovanova, X. He, A. Mahadevan, J. Ahn, O. Stanoi, J. K. Brynildsen, N. Cooper, E. J. Cornblath, L. Parkes, P. J. Mucha, K. N. Ochsner, D. M. Lydon-Staley, E. B. Falk, and D. S. Bassett, "Mindful Attention Promotes Control of Brain Network Dynamics for Self-Regulation and Discontinues the Past from the Present," Proceedings of the National Academy of Sciences, vol. 120, no. 2, p. e2201074119, Jan. 2023.

[9] X. He, L. Caciagli, L. Parkes, J. Stiso, T. M. Karrer, J. Z. Kim, Z. Lu, T. Menara, F. Pasqualetti, M. R. Sperling, J. I. Tracy, and D. S. Bassett, "Uncovering the Biological Basis of Control Energy: Structural and Metabolic Correlates of Energy Inefficiency in Temporal Lobe Epilepsy," Science Advances, vol. 8, no. 45, p. eabn2293, Nov. 2022.

[10] L. Tang, P. Zhao, C. Pan, Y. Song, J. Zheng, R. Zhu, F. Wang, and Y. Tang, "Epigenetic molecular underpinnings of brain structural functional connectivity decoupling in patients with major depressive disorder," Journal of Affective Disorders, vol. 363, pp. 249–257, 2024.

[11] C. Pan, Y. Ma, L. Wang, Y. Zhang, F. Wang, and X. Zhang, "From connectivity to controllability: Unraveling the brain biomarkers of major depressive disorder," Brain Sciences, vol. 14, no. 5, p. 509, 2024.

[12] H. Guo, Y. Xiao, D. Sun, J. Yang, J. Wang, H. Wang, C. Pan, C. Li, P. Zhao, Y. Zhang et al., "Early-stage repetitive transcranial magnetic stimulation altered posterior–anterior cerebrum effective connectivity in methylazoxymethanol acetate rats," Frontiers in Neuroscience, vol. 15, p. 652715, 2021.

[13] Y. Chen, P. Zhao, C. Pan, M. Chang, X. Zhang, J. Duan, Y. Wei, Y. Tang, and F. Wang, "State-and trait-related dysfunctions in bipolar disorder across different mood states: a graph theory study," Journal of Psychiatry and Neuroscience, vol. 49, no. 1, pp. E11–E22, 2024.

[14] C. Pan, Q. Zhang, Y. Zhu, S. Kong, J. Liu, C. Zhang, F. Wang, and X. Zhang, "Module control of network analysis in psychopathology," iScience, vol. 27, no. 7, 2024.

[15] M. Bardoscia, P. Barucca, S. Battiston, F. Caccioli, G. Cimini, D. Gar laschelli, F. Saracco, T. Squartini, and G. Caldarelli, "The Physics of Financial Networks," Nature Reviews Physics, vol. 3, no. 7, pp. 490 507, Jul. 2021, publisher: Nature Publishing Group.

[16] L. F. Seoane, "Games in Rigged Economies," Physical Review X, vol. 11, no. 3, p. 031058, Sep. 2021.

[17] A. Ciullo, E. Strobl, S. Meiler, O. Martius, and D. N. Bresch, "Increas ing Countries' Financial Resilience through Global Catastrophe Risk Pooling," Nature Communications, vol. 14, no. 1, p. 922, Feb. 2023.

[18] Y.-Y. Liu, J.-J. Slotine, and A.-L. Barab´asi, "Controllability of Complex Networks," Nature, vol. 473, no. 7346, pp. 167–173, May 2011.

[19] R. M. D'Souza, M. di Bernardo, and Y.-Y. Liu, "Controlling Complex Networks with Complex Nodes," Nature Reviews Physics, vol. 5, no. 4, pp. 250–262, Apr. 2023, publisher: Nature Publishing Group.

[20] G. Baggio, D. S. Bassett, and F. Pasqualetti, "Data-Driven Control of Complex Networks," Nature Communications, vol. 12, no. 1, p. 1429, Mar. 2021, publisher: Nature Publishing Group.

[21] Y.-Y. Liu and A.-L. Barab´asi, "Control Principles of Complex Systems," Reviews of Modern Physics, vol. 88, no. 3, p. 035006, Sep. 2016.

[22] J. Ruths and D. Ruths, "Control Profiles of Complex Networks," Science, vol. 343, no. 6177, pp. 1373–1376, Mar. 2014, publisher: American Association for the Advancement of Science.

[23] C. Campbell, J. Ruths, D. Ruths, K. Shea, and R. Albert, "Topological Constraints on Network Control Profiles," Scientific Reports, vol. 5, no. 1, p. 18693, Dec. 2015.

[24] L. G. Valiant, "The complexity of computing the permanent," Theoret ical computer science, vol. 8, no. 2, pp. 189–201, 1979.

[25] X. Zhang, J. Han, and W. Zhang, "An Efficient Algorithm for Finding All Possible Input Nodes for Controlling Complex Networks," Scientific Reports, vol. 7, no. 1, p. 10677, Sep. 2017, publisher: Nature Publishing Group.

[26] X. Zhang, C. Pan, and W. Zhang, "Control hubs of complex networks and a polynomial-time identification algorithm," 2022. [Online]. Available: https://arxiv.org/abs/2206.01188

[27] X. Zhang, H. Wang, and T. Lv, "Efficient Target Control of Complex Networks Based on Preferential Matching," PLOS ONE, vol. 12, no. 4, p. e0175375, Apr. 2017.

[28] X. Zhang, T. Lv, X. Yang, and B. Zhang, “Structural Controllability of Complex Networks Based on Preferential Matching,” PLoS ONE, vol. 9, no. 11, p. e112039, Nov. 2014.
[29] X. Zhang, Y. Zhu, and Y. Zhao, “Altering Control Modes of Complex Networks by Reversing Edges,” Physica A: Statistical Mechanics and its Applications, vol. 561, p. 125249, Jan. 2021.
[30] X. Zhang and Q. Li, “Altering Control Modes of Complex Networks Based on Edge Removal,” Physica A: Statistical Mechanics and its Applications, vol. 516, pp. 185–193, Feb. 2019.
[31] A. Li, S. P. Cornelius, Y.-Y. Liu, L. Wang, and A.-L. Barab´asi, “The Fundamental Advantages of Temporal Networks,” Science, vol. 358, no. 6366, pp. 1042–1046, Nov. 2017.
[32] Y. Cui, S. He, M. Wu, C. Zhou, and J. Chen, “Improving the Con trollability of Complex Networks by Temporal Segmentation,” IEEE Transactions on Network Science and Engineering, vol. 7, no. 4, pp. 2765–2774, Oct. 2020.
[33] M. P´osfai and P. H¨ovel, “Structural Controllability of Temporal Net works,” New Journal of Physics, vol. 16, no. 12, p. 123055, Dec. 2014.
[34] B. Ravandi, F. Mili, and J. A. Springer, “Identifying and Using Driver Nodes in Temporal Networks,” Journal of Complex Networks, vol. 7, no. 5, pp. 720–748, Oct. 2019.
[35] P. Yao, B.-Y. Hou, Y.-J. Pan, and X. Li, “Structural Controllability of Temporal Networks with a Single Switching Controller,” PLOS ONE, vol. 12, no. 1, p. e0170584, Jan. 2017.
[36] Y. Pan and X. Li, “Structural Controllability and Controlling Centrality of Temporal Networks,” PLoS ONE, vol. 9, no. 4, p. e94998, Apr. 2014.
[37] T. Jia, Y.-Y. Liu, E. Cs´oka, M. P´osfai, J.-J. Slotine, and A.-L. Barab´asi, “Emergence of bimodality in controlling complex networks,” Nature communications, vol. 4, no. 1, pp. 1–6, 2013.
[38] J.-c. Tu, H.-q. Lu, T.-m. Lu, Z.-q. Xie, L. Lu, and L. Wei, “A graphical criterion for the controllability in temporal networks,” Physica A: Statistical Mechanics and its Applications, vol. 646, p. 129906, 2024.
[39] Y. Qin and K. Yan, “A novel recovery controllability method on temporal networks via temporal lost link prediction,” Journal of Complex Networks, vol. 12, no. 6, p. cnae042, 2024.
[40] M. V. Srighakollapu, R. K. Kalaimani, and R. Pasumarthy, “Optimizing driver nodes for structural controllability of temporal networks,” IEEE Transactions on Control of Network Systems, vol. 9, no. 1, pp. 380–389, 2021.
[41] B. Ravandi, F. Mili, and J. A. Springer, “Identifying and using driver nodes in temporal networks,” Journal of Complex Networks, vol. 7, no. 5, pp. 720–748, 2019.
[42] T. Qin, G. Duan, and A. Li, “Detecting the driver nodes of temporal networks,” New Journal of Physics, vol. 25, no. 8, p. 083031, 2023.
[43] F. Li, “Improving the efficiency of network controllability processes on temporal networks,” Journal of King Saud University-Computer and Information Sciences, vol. 36, no. 3, p. 101976, 2024.
[44] H. Kwakernaak and R. Sivan, Linear optimal control systems. Wiley interscience New York, 1972, vol. 1.
[45] A. Li, S. P. Cornelius, Y.-Y. Liu, L. Wang, and A.-L. Barab´asi, “The fundamental advantages of temporal networks,” Science, vol. 358, no. 6366, pp. 1042–1046, 2017.
[46] L. E. C. Rocha, N. Masuda, and P. Holme, “Sampling of Temporal Networks: Methods and Biases,” Physical Review E, vol. 96, no. 5, p. 052302, Nov. 2017.
[47] E. Almaas, B. Kov´acs, T. Vicsek, Z. N. Oltvai, and A.-L. Barab´asi, “Global Organization of Metabolic Fluxes in the Bacterium Escherichia Coli,” Nature, vol. 427, no. 6977, pp. 839–843, Feb. 2004.
[48] P. Holme and J. Saram¨aki, “Temporal Networks,” Physics Reports, vol. 519, no. 3, pp. 97–125, Oct. 2012.
[49] R. E. Kalman, “Mathematical Description of Linear Dynamical Sys tems,” Journal of the Society for Industrial and Applied Mathematics Series A Control, vol. 1, no. 2, pp. 152–192, Jan. 1963, publisher: Society for Industrial and Applied Mathematics.
[50] J. E. Hopcroft and R. M. Karp, “An \n^{5/2} \ Algorithm for Maximum Matchings in Bipartite Graphs,” SIAM Journal on Computing, vol. 2, no. 4, pp. 225–231, Dec. 1973.
[51] L. Zdeborov´a and M. M´ezard, “The Number of Matchings in Random Graphs,” Journal of Statistical Mechanics: Theory and Experiment, vol. 2006, no. 05, pp. P05003–P05003, May 2006.
[52] T. Jia and A.-L. Barab´asi, “Control Capacity and A Random Sampling Method in Exploring Controllability of Complex Networks,” Scientific Reports, vol. 3, no. 1, p. 2354, Aug. 2013.
[53] A.-L. Barab´asi, “Linked: The new science of networks,” 2003.
[54] R. Albert and A.-L. Barab´asi, “Statistical Mechanics of Complex Net works,” Reviews of Modern Physics, vol. 74, no. 1, pp. 47–97, Jan. 2002.
[55] V. Mart´ ınez, F. Berzal, and J.-C. Cubero, “A Survey of Link Prediction in Complex Networks,” ACM Computing Surveys, vol. 49, no. 4, pp. 69:1–69:33, Dec. 2016.
[56] A.-L. Barab´asi and R. Albert, “Emergence of Scaling in Random Networks,” Science, vol. 286, no. 5439, pp. 509–512, Oct. 1999.
[57] P. ERDdS and A. R&wi, “On Random Graphs I,” Publ. math. debrecen, vol. 6, no. 290-297, p. 18, 1959.
[58] P. Panzarasa, T. Opsahl, and K. M. Carley, “Patterns and Dynamics of Users’ Behavior and Interaction: Network Analysis of an Online Community,” Journal of the American Society for Information Science and Technology, vol. 60, no. 5, pp. 911–932, May 2009.
[59] A. Paranjape, A. R. Benson, and J. Leskovec, “Motifs in Temporal Networks,” in Proceedings of the Tenth ACM International Conference on Web Search and Data Mining, ser. WSDM ’17. New York, NY, USA: Association for Computing Machinery, Feb. 2017, pp. 601–610.
[60] R. Rossi and N. Ahmed, “The Network Data Repository with Interactive Graph Analytics and Visualization,” Proceedings of the AAAI Conference on Artificial Intelligence, vol. 29, no. 1, Mar. 2015.
[61] P. Jaccard, “´ Etude comparative de la distribution florale dans une portion des alpes et des jura,” Bull Soc Vaudoise Sci Nat, vol. 37, pp. 547–579, 1901.
[62] K. Faust, “Centrality in affiliation networks,” Social networks, vol. 19, no. 2, pp. 157–191, 1997.
[63] L. Page, S. Brin, R. Motwani, and T. Winograd, “The pagerank citation ranking: Bringing order to the web.” Stanford infolab, Tech. Rep., 1999.